\documentclass[sigconf]{acmart}
\AtBeginDocument{%
  }

\usepackage{amsmath,amssymb,amsfonts}
\usepackage{algorithmic}
\usepackage{graphicx}
\usepackage{textcomp}
\usepackage{xcolor}
\usepackage{xspace}
\usepackage{enumerate}
\usepackage{url}
\usepackage{booktabs}
\usepackage{graphicx}
\usepackage{subfigure} 
\usepackage{bm}
\usepackage{multirow}
\usepackage{array}
\usepackage{makecell}
\usepackage{tabularx}
\usepackage{stfloats}
\usepackage{makecell}
\usepackage{pifont}
\usepackage{enumitem}
\usepackage{scalerel}

\setcopyright{acmlicensed}
\copyrightyear{2026}
\acmYear{2026}
\acmISBN{978-1-4503-XXXX-X/2018/06}
\acmDOI{XXXXXXX.XXXXXXX}
\acmConference[MobiSys '26]{The 24th ACM International Conference on Mobile Systems, Applications, and Services}{June 21--25, 2026}{Cambridge, UK}
\acmBooktitle{The 24th ACM International Conference on Mobile Systems, Applications, and Services (MobiSys '26), June 21--25, 2026, Cambridge, UK}

\begin{document}

\newcommand{\systemname}{BSense\xspace}
\newcommand{\SYSTEMNAME}{BSENSE\xspace}

\title{Needle in a Haystack: Tracking UAVs from Massive Noise in Real-World 5G-A Base Station Data}


\renewcommand\footnotemark{}

\author{
  Chengzhen Meng${^\dag} {^*}$, Chenming He${^\dag} {^*}$, Yidong Jiang$^\dag$, Xiaoran Fan$^\ddag$, Dequan Wang$^\dag$,\\
  Lingyu Wang$^\dag$, Jianmin Ji$^\dag$, Yanyong Zhang\textsuperscript{$\dag \diamondsuit$}\\
  \vspace{3pt}
  \large $^\dag$University of Science and Technology of China, $^\ddag$Independent Researcher,\\
  $^\diamondsuit$Institute of Artificial Intelligence, Hefei Comprehensive National Science Center\\
  \vspace{3pt}
  \{czmeng, hechenming, yyydddd, wdq15588, lywang19\}@mail.ustc.edu.cn,\\
  \vspace{-2pt}
  gunanjiluzhe@gmail.com, \{jianmin, yanyongz\}@ustc.edu.cn
  }
\authornote{Equal contribution.}


\renewcommand{\shortauthors}{Chengzhen Meng, Chenming He, et al.}

\begin{abstract}
The potential usage of UAVs in daily life has made monitoring them essential.
However, existing systems for monitoring UAVs typically rely on cameras, LiDARs, or radars, whose limited sensing range or high deployment cost hinder large-scale adoption.
In response, we develop \systemname, the first system that tracks UAVs by leveraging point clouds from commercial 5G-A base stations.
The key challenge lies in the dominant number of noise points that closely resemble true UAV points, resulting in a noise-to-UAV ratio over 100:1.
Therefore, identifying UAVs from the raw point clouds is like finding a needle in a haystack.
To overcome this, we propose a layered framework that filters noise at the point, object, and trajectory levels.
At the raw point level, we observe that noise points from different spatial regions exhibit distinguishable and consistent signal fingerprints, which we can model to identify and remove them.
At the object level, we design spatial and velocity consistency checks to identify false objects, and further compute confidence scores by aggregating these checks over multiple frames for more reliable discrimination.
At the final trajectory level, we propose a Transformer-based network that captures multi-frame motion patterns to filter the few remaining false trajectories.

We evaluated \systemname on a commercial 5G-A base station deployed in an urban environment. The UAV was instructed to fly along 25 distinct trajectories across 54 cases over 7 days, yielding 155 minutes of data with more than 14,000 frames. On this dataset, our system reduces the number of false detections from an average of 168.05 per frame to 0.04, achieving an average F1 score of 95.56\% and a mean localization error of 4.9~m at ranges up to 1,000~m.
\end{abstract}





\maketitle

\section{Introduction}
\label{sec:intro}


Unmanned aerial vehicles (UAVs) are becoming increasingly prevalent, with applications ranging from aerial transportation to infrastructure inspection~\cite{zhao2024few, alqudsi2025uav, zhou2025unmanned}.
They are also emerging as integral components of smart city infrastructure, playing vital roles in traffic monitoring, emergency response, and environmental surveillance~\cite{mohamed2020unmanned, sharma2022uav, chen2024uitde}.
Consequently, accurate wide-area tracking of UAVs is essential for these applications, ensuring safe flight navigation and reliable airspace intrusion detection.

Prior studies have explored the use of cameras, LiDARs, and radars for UAV perception~\cite{wang2021deep, wang2024survey, zhao2022vision, zhang2023mmhawkeye, abir2025detection}. 
However, cameras suffer from limited sensing range~\cite{li2024farfusion} and high sensitivity to lighting conditions~\cite{rezaei2015robust}, while LiDARs and radars are costly and impractical for large-scale deployment~\cite{poitevin2017challenges}.
Consequently, their limited coverage and high deployment cost make them inadequate for monitoring UAVs that may appear at arbitrary locations in urban environments. 
To overcome these limitations, we explore leveraging existing base station infrastructure to develop a low-cost UAV detection and tracking system with wide-area coverage.

Recently, 5G has been advancing toward 5G-Advanced (5G-A), which incorporates Integrated Sensing and Communication (ISAC) technology to enhance base station functionality~\cite{chen2024resilient, wei2024multiple, luo2025isac, liu2025carrier}.
In addition to providing communication services, 5G-A base stations also offer sensing abilities~\cite{wei2023integrated, lin2024integrated}.
This enables low-cost, wide-area sensing of UAVs by leveraging existing base station infrastructure without requiring additional sensing equipment~\cite{wu2021comprehensive}.

There have been previous studies~\cite{zhou2024joint, lu2024integrated, motie2024self, elfiatoure2025multiple, li2025joint, yang2025hierarchical} that explore UAV tracking using 5G-A base stations.
However, most of them are limited to theoretical analysis or idealized simulations, which fail to reflect the complexity of real-world physical environments.
The only reported real-world deployment is by Lu et al.~\cite{lu2024intergrated}, which focuses on maritime vessel detection in open environments where large targets are relatively easy to detect. As a result, the feasibility of wide-area UAV tracking in complex urban environments remains largely unexplored.
To move beyond these simplified settings, we investigate the feasibility of wide-area UAV tracking using point clouds from a commercial off-the-shelf (COTS) 5G-A base station deployed in an urban environment.
To the best of our knowledge, this is the first work to reveal and address the key challenge that emerges in such real-world deployment: \emph{massive and persistent false points which closely resemble true UAV points.}

\begin{figure}
    \centering
    \setlength{\abovecaptionskip}{1mm}
    \includegraphics[width=0.99\linewidth]{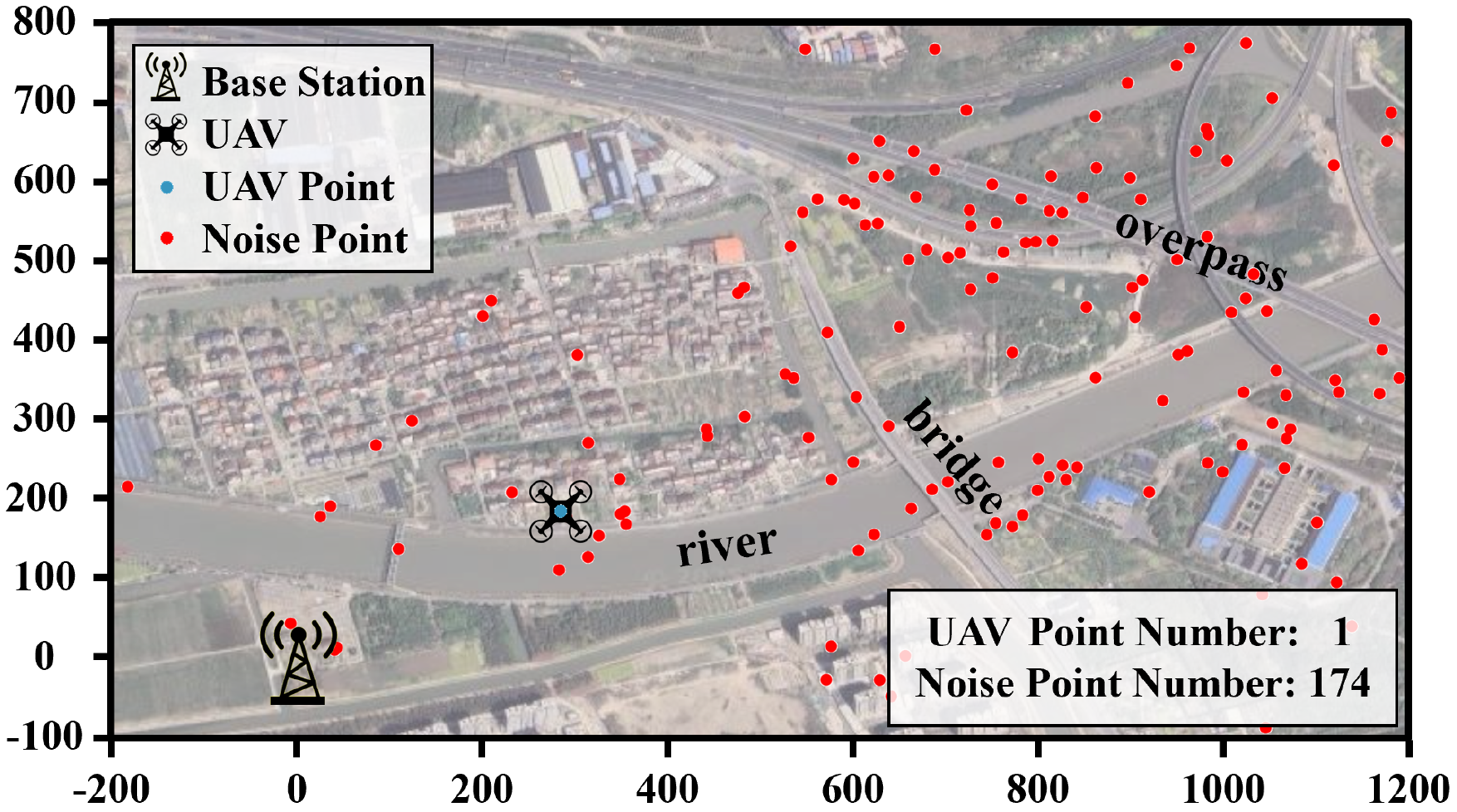}
    \caption{A single frame of real-world data. \textmd{The number of noise points exceeds UAV points by a ratio of 174:1.}}
    \label{fig: problem}
    \vspace{-2mm}
\end{figure}

\begin{figure}
    \centering
    \setlength{\abovecaptionskip}{1mm}
    \includegraphics[width=0.99\linewidth]{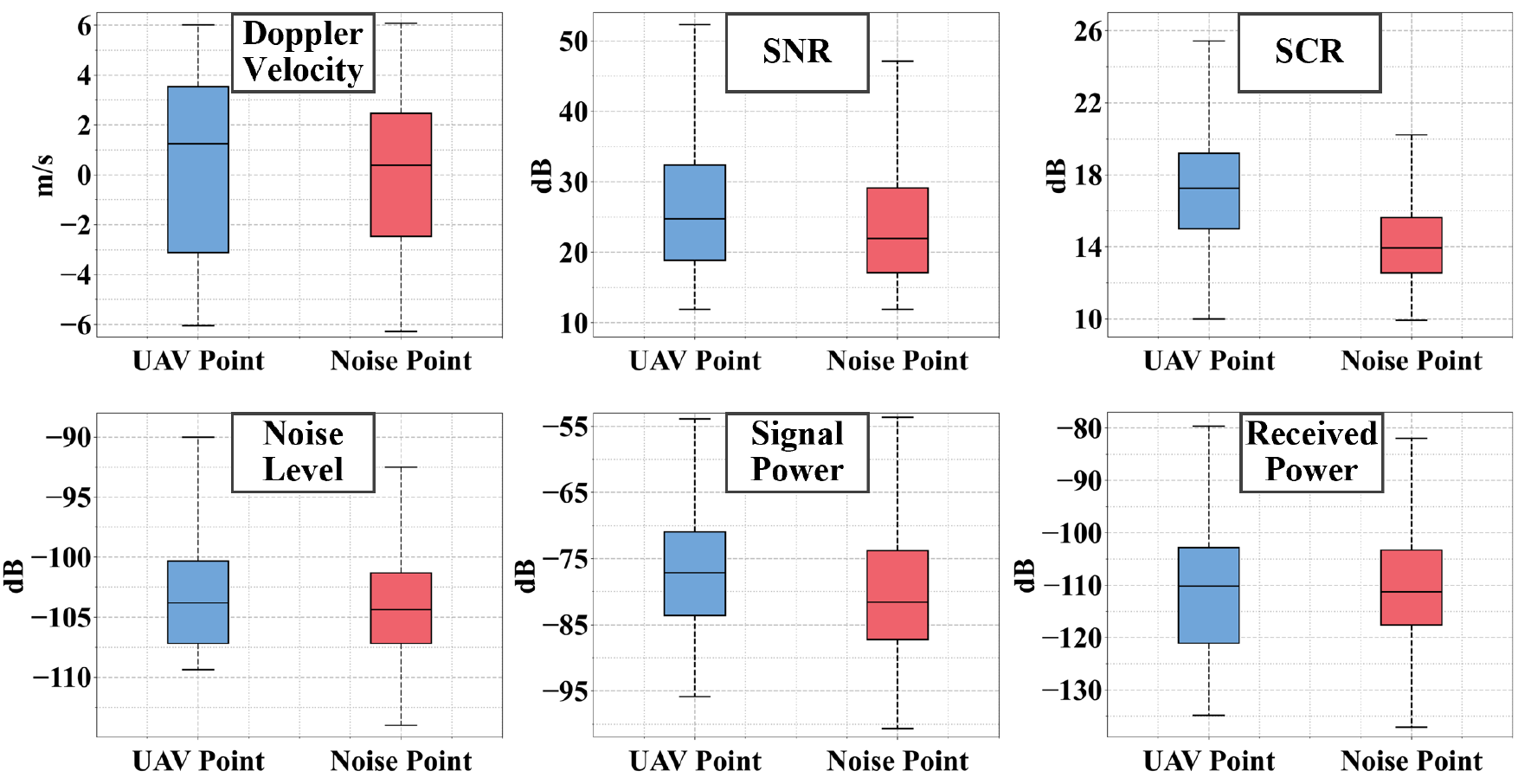}
    \caption{The value ranges of point features for UAV and noise exhibit substantial overlap.} 
    \label{fig: challenge_1}
    \vspace{-3mm}
\end{figure}

To substantiate this claim, Figure~\ref{fig: problem} shows a single frame from our dataset.
The data were collected on a Huawei 5G-A base station~\cite{core_network, Huawei2024CloudCore} deployed in Shanghai, which generates point clouds through its internal signal-processing pipeline.
For clearer visualization, we project the point cloud onto the corresponding satellite map. 
Notably, unlike prior systems that focus on UAVs within short distances (typically under 150 m)~\cite{lu2024integrated, yang2025hierarchical}, the COTS base station we used provides a sensing range of up to 1,000 m.
However, at such long ranges, UAV reflections become extremely weak and are difficult to distinguish from noise.
Thus, identifying UAVs from raw point clouds is like finding a needle in a haystack. 

Although signal-level noise suppression is applied before point cloud generation, a substantial number of noise points still persist.
In this frame (Figure~\ref{fig: problem}), only a single point corresponds to the UAV, while 174 points are noise---resulting in a two-order-of-magnitude imbalance. 
Through theoretical analysis and experimental observations, we attribute these noise points to three sources: 
(1) \textit{Background clutter} from stationary structures (e.g., buildings and trees). Notably, small physical vibrations and Doppler leakage can make their returns exhibit non-zero velocities and thus difficult to filter~\cite{liu2025key}.
(2) \textit{Multi-path ghosts} caused by multiple reflections between moving objects (e.g., vehicles) and reflective surfaces (e.g., building facades).
(3) \textit{Sidelobe-induced ghosts} caused by sidelobe returns incorrectly interpreted as main-lobe signals.
Therefore, our task is to accurately distinguish UAVs from these abundant false detections, which has not been investigated in prior works. 

\begin{figure}
    \centering
    \setlength{\abovecaptionskip}{1mm}
    \includegraphics[width=0.99\linewidth]{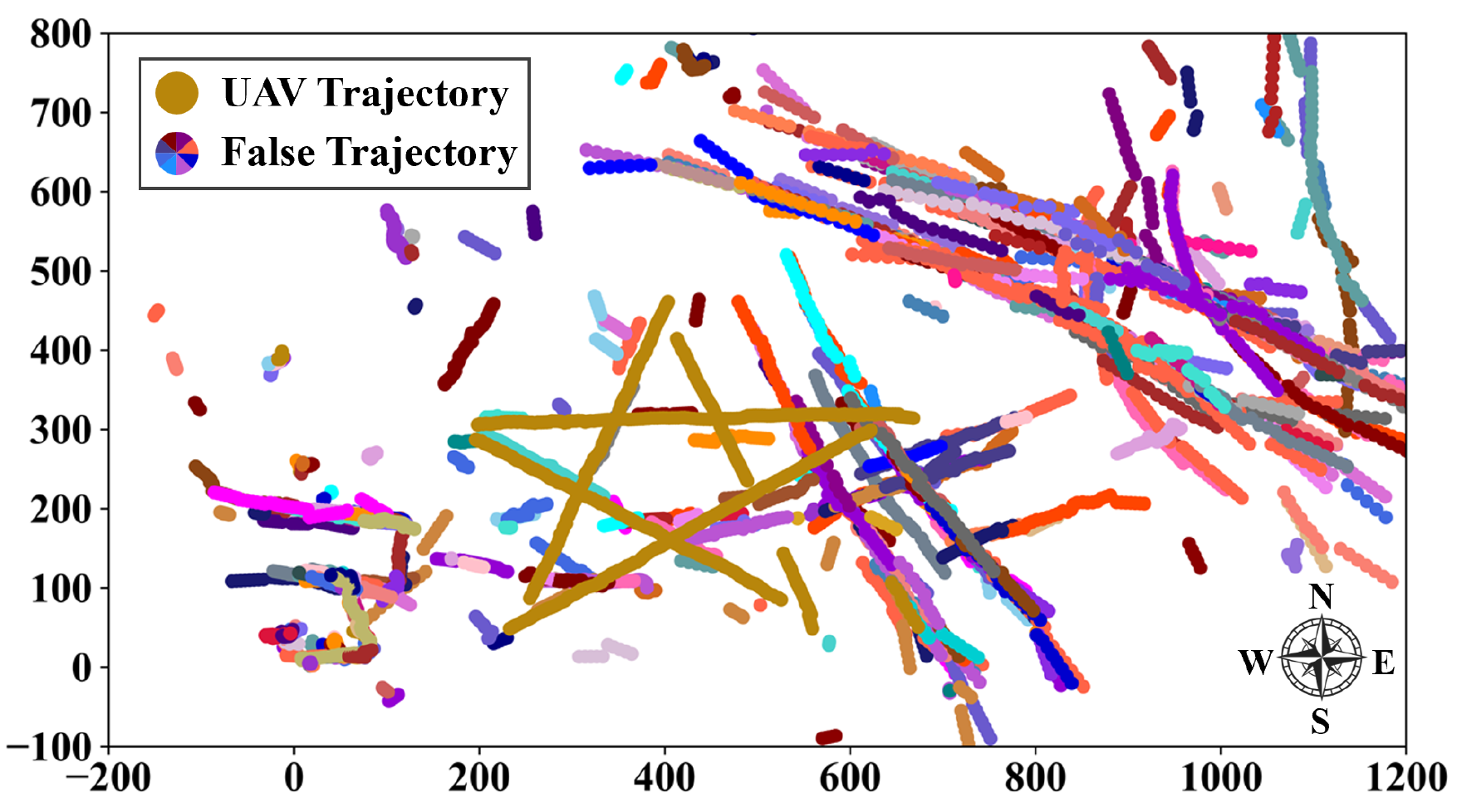}
    \caption{Persistent noise points give rise to massive long-term false trajectories.}
    \label{fig: challenge_2}
    \vspace{-3mm}
\end{figure}

However, this isn't easy because the false detections are persistent and significantly similar to UAV targets:
\begin{itemize}
\vspace{-2pt}
    \item \textit{Feature Overlap Between UAV and Noise Points.} 
    As illustrated in Figure~\ref{fig: challenge_1}, a statistical analysis of point cloud data from the COTS base station, including Doppler velocity, SNR, SCR, and other signal metrics, shows substantial overlap between the feature distributions of UAV points and noise points across all dimensions. This strong similarity eliminates any clear decision boundary, rendering traditional threshold-based filtering methods ineffective and leaving noise points difficult to distinguish from true UAV points.
    \item \textit{Long-term False Trajectories from Persistent Noise Points.} 
    Beyond the similarity in point-level features, many noise points persist over long durations, forming stable trajectories that closely mimic those of UAVs. As shown in Figure~\ref{fig: challenge_2}, we cluster and track the raw points and visualize the trajectories in different colors. The true UAV trajectory appears as a dark gold \ding{73}, yet massive false trajectories also arise, some extending for hundreds of meters. 
    Critically, these false trajectories often occur in plausible UAV locations, making accurate discrimination particularly challenging. 
\vspace{-2pt}
\end{itemize}

A straightforward approach is to use deep learning to classify and filter out noise points and false trajectories. However, this approach is challenged by the overwhelming number of noise points and their high similarity to true UAV points, which makes accurate discrimination difficult. As shown in Section~\ref{sec:baseline-compare}, even state-of-the-art learning methods fail to eliminate the noise effectively.

To overcome these challenges, we propose \systemname, which achieves wide-area detection and tracking of UAVs using a COTS 5G-A base station.
\systemname employs a layered framework that filters noise at the point, object, and trajectory levels while preserving true UAV targets.
At the point level, our key insight is that the background noise is strongly correlated with the surrounding environment and thus exhibits statistical patterns within local spatial regions. 
After partitioning the 3D sensing space into cubes, we observe that the noise points within each cube form a distinctive ``fingerprint'' consisting of multiple signal metrics.
We therefore model each cube's noise fingerprint as a Gaussian distribution and estimate its mean vector and covariance matrix. 
Raw points are then evaluated by their similarity to the corresponding fingerprint, and those with a high degree of similarity are filtered out.
Notably, only 10 minutes of data are sufficient for modeling, and due to the slow temporal variation, the model needs to be updated just once per day.

After point-level filtering, we cluster the points into individual objects and further eliminate false objects.
Specifically, we check object consistency between two consecutive frames based on intuitive motion principles.
(1) \emph{Spatial consistency:} True UAVs exhibit continuous spatial motion. Thus, a true object should have a corresponding object within a spatial neighborhood in the previous frame; if no such object exists, it is regarded as a false object.
(2) \emph{Velocity consistency:} A true object's Doppler velocity measured by the base station should align with the radial velocity computed from the inter-frame spatial displacement; otherwise, it is regarded as a false object.
However, consistency checks based on only two consecutive frames can be unreliable due to measurement noise. To address this, we aggregate these binary outcomes over time using an exponentially decaying weighting strategy, yielding spatial- and velocity-based confidence scores that emphasize more recent observations.
This design produces robust indicators of object validity, enabling the system to reliably filter out objects that lack stable spatial or velocity behavior.

\begin{figure*}
    \centering
    \setlength{\abovecaptionskip}{1mm}
    \includegraphics[width=0.99\linewidth]{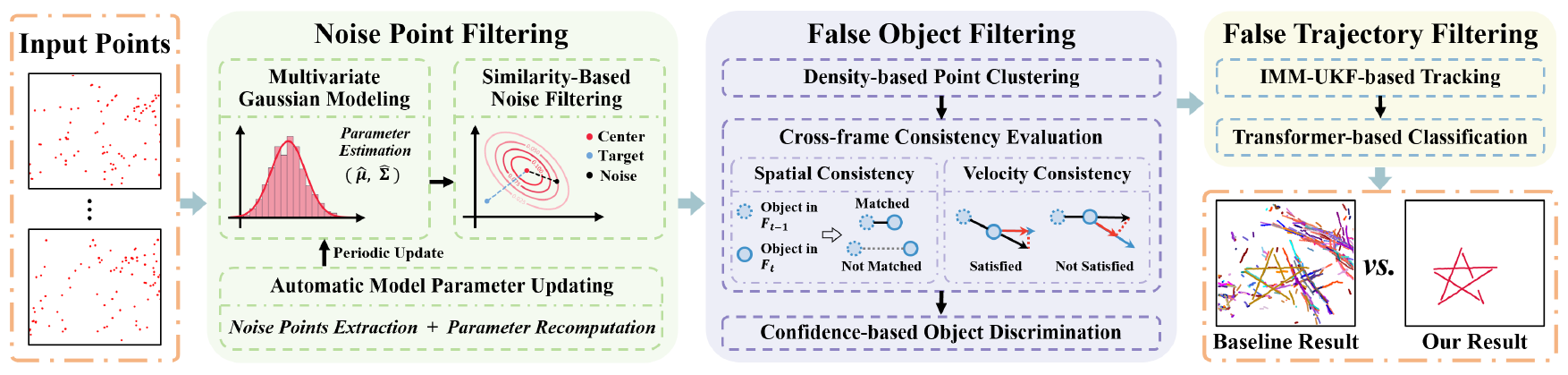}
    \caption{\systemname adopts a layered framework consisting of (1) noise point filtering, (2) false object filtering, and (3) false trajectory filtering, which progressively suppress false detections to yield clean UAV trajectories.}
    \label{fig: overview}
    \vspace{-2mm}
\end{figure*}

Finally, objects are tracked into individual trajectories; however, some false ones may still persist.
To address this, we propose TrajFormer, a lightweight trajectory classification network inspired by the Transformer architecture~\cite{vaswani2017attention}, which leverages spatial motion and signal features across multiple frames to filter false trajectories. 
By capturing long-range temporal dependencies and motion dynamics with its encoder, and aligning the current observation with historical context through its decoder, TrajFormer effectively distinguishes true trajectories from false ones.

In summary, we make the following contributions:

$\bullet$ 
We propose \systemname, the first system that achieves accurate wide-area UAV detection and tracking using a commercial 5G-A base station in a real-world deployment.
A demonstration video is available at https://youtu.be/Bx668ORkAq4.

$\bullet$
We propose a layered framework that filters noise at the point, object, and trajectory levels while preserving true UAV targets.
First, we introduce an adaptive noise fingerprint modeling method to remove points similar to the noise.
Next, we design spatial and velocity consistency checks to distinguish true UAV targets from false objects.
Finally, we employ a lightweight classification Transformer-based network to eliminate the residual false trajectories.

$\bullet$
We utilized a COTS 5G-A base station deployed in an urban environment and commanded the UAV to follow diverse flight trajectories ($-$,\; $\bigcirc$,\; $\text{S}$,\; $\infty$,\; $\square$,\; $\scalerel*{\diamond}{\square}$,\; $\text{M}$,\; \ding{73}). 
Across 7 days, we collected 54 flight cases, yielding approximately 155 minutes of valid data (over 14,000 frames).
We evaluate \systemname on this dataset and achieve an average F1 score of 95.56\%, with 96.73\% precision, 94.41\% recall, and a mean localization error of 4.9~m, demonstrating its accuracy and robustness in real-world urban conditions.

\section{Background}

In this paper, we utilize a COTS Huawei 5G-A base station (4.9–4.91 GHz, 128 channels)~\cite{core_network, Huawei2024CloudCore}.
The base station adopts a hybrid OFDM–LFM waveform~\cite{5G-A_huawei}, where OFDM supports short-range coverage while LFM enables long-range sensing. 
Internally, the base station employs a customized radar-style signal processing pipeline~\cite{wei2023integrated, imt_2020} to estimate range, Doppler, and angle, and directly outputs point clouds. 
In this work, we do not access or modify the underlying signal processing modules of the base station. Instead, we directly use the generated point clouds as input and focus on application-layer perception tasks built upon them.

Based on the point clouds generated by the base station, our goal is to achieve accurate UAV detection and tracking while effectively mitigating false detections.
Specifically, our system takes as input a point cloud $P$, where each point $p \in P$ encodes spatial coordinates, Doppler velocity, and a set of signal quality metrics, including signal-to-noise ratio (SNR), signal-to-clutter ratio (SCR), noise level, signal power, and received power:
\begin{equation}
\small
p = \left( x, \; y, \; z, \; v_d, \; snr, \; scr, \; nl, \; sp, \; rp \right), \quad p \in P.
\end{equation}
The output of our system is the estimated position, velocity, and identity of each tracked UAV, represented as:
\begin{equation}
\small
u = \left( x, \; y, \; z, \; v_x, \; v_y, \; v_z, \; id \right), \quad u \in U.
\end{equation}

\section{\systemname Design}


\subsection{System Overview}

Figure~\ref{fig: overview} presents an overview of \systemname, a layered framework operating at the point, object, and trajectory levels.
At the point level, the module models and automatically updates the noise fingerprint, filtering points that are similar to the noise fingerprint.
At the object level, the module clusters points into objects, performs spatial and velocity consistency checks, and computes confidence scores by aggregating these checks over multiple frames to distinguish true targets from false objects.
At the trajectory level, the module tracks objects and employs a lightweight Transformer-based classification network to eliminate false trajectories.

\subsection{Noise Point Filtering through Noise Fingerprint Modeling}
\label{sec: noise_point_filtering}
We start noise filtering at the point cloud level, targeting persistent background noise. 
These are primarily caused by direct reflections from stationary structures, such as buildings, terrain, and trees. 
However, due to environmental vibrations and Doppler leakage~\cite{liu2025key}, these points may exhibit non-zero Doppler velocities (up to 3.2~m/s) and high signal strength, making them difficult to distinguish from UAV points based on raw point features directly.

To address this, we shift our focus from individual point-level features to the statistical patterns they form within local spatial regions. 
Our key idea is that background noise arises from echo signals reflected by the surrounding environment. 
Therefore, the signal metrics of noise points within each local region should follow a characteristic and consistent distribution, which we refer to as the \textbf{noise fingerprint}. 
Building on this insight, we first analyze and confirm that the noise fingerprint can be effectively characterized by a Gaussian distribution. 
We then estimate the distribution parameters across multiple frames and filter points based on their statistical deviation from the fitted distribution. 

\subsubsection{Analysis of Noise Fingerprint}

We find that signal metrics of noise points within local spatial regions can be effectively modeled as a Gaussian distribution. 
This conclusion is supported by both theoretical reasoning and empirical evidence, as detailed below.

From a theoretical perspective, this assumption is supported by two key observations.
First, by the Central Limit Theorem (CLT), the sum of a large number of independent random variables converges to a Gaussian distribution. 
In our scenario, noise signals in a local region result from the aggregation of reflections from numerous small, independent scatterers.
Second, within a sufficiently small spatial region, the physical properties of the environment are typically homogeneous. This spatial consistency leads to stable scattering behavior, supporting the convergence of aggregated signal metrics toward a Gaussian distribution.

To empirically validate this hypothesis, we collect 10 minutes of noise data (900 frames) and accumulate the corresponding point clouds. 
The 3D space is partitioned into cubes with an edge length of 40~m, each containing a localized set of points. 
For each cube, we apply the Henze-Zirkler test~\cite{H-Ztest} to assess whether the five signal metrics---SNR, SCR, noise level, signal power, and received power---jointly follow a multivariate Gaussian distribution. 
Results show that over 75\% of cubes satisfy the Gaussianity assumption, confirming the feasibility of statistical modeling. 
For the remaining cubes that fail the test, we observe that they often contain a higher proportion of complex ghost points, likely caused by multi-path effects or sidelobe interference, which leads to non-Gaussian signal patterns. 
These cubes are excluded from Gaussian modeling, and their points are handled in later stages, where BSense’s object-level (Section~\ref{sec: fake_object_filtering}) and trajectory-level (Section~\ref{sec: fake_trajectory_filtering}) modules further identify and remove such ghost points.

In the following sections, we detail the procedures for Gaussian modeling, noise filtering, and parameter updating.


\subsubsection{Multivariate Gaussian Modeling.}
\label{sec: gaussian_model}

\begin{figure}
    \centering
    \setlength{\abovecaptionskip}{1mm}
    \includegraphics[width=0.99\linewidth]{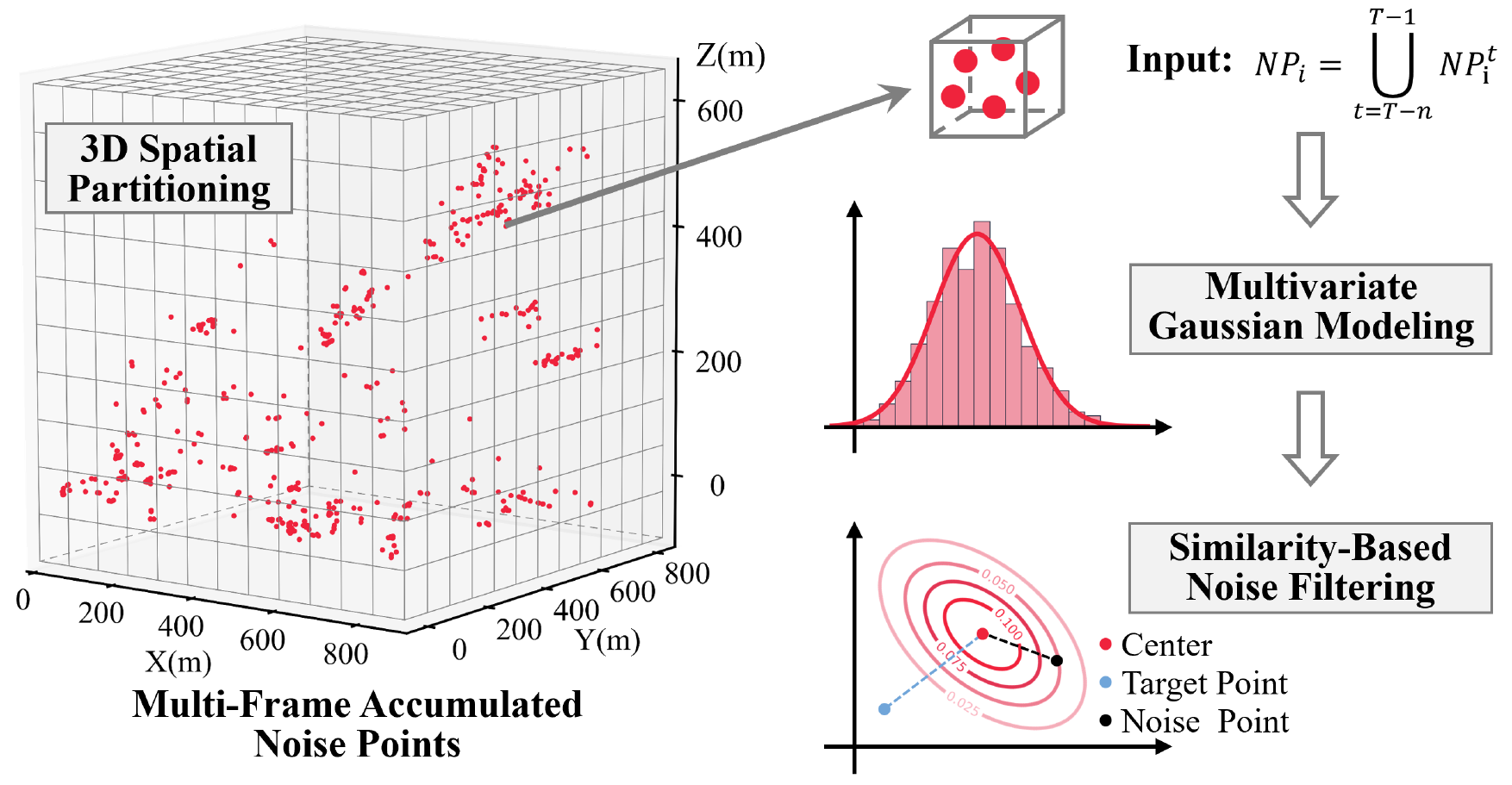}
    \caption{Noise point filtering through fingerprint modeling. \textmd{The 3D space is partitioned into cubes, and the noise within each cube is modeled using a Gaussian distribution. Points are then filtered based on their similarity to the corresponding distribution.}}
    \label{fig: cube}
    \vspace{-3mm}
\end{figure}

As illustrated in Figure~\ref{fig: cube}, we partition the 3D space into cubes (each with an edge length of 40~m) and model the noise distribution within each cube using a multivariate Gaussian.
Due to space limitations, the rationale behind the cube size selection is discussed in Appendix~\ref{appendix-b}.
To construct the initial noise model, we first collect point cloud data over a period during which no UAVs are present in the sensing area, denoted as $NP$. This ensures that all points in $NP$ can be regarded as noise. For each cube $i$, we define $NP_i$ as the subset of noise points located within it:
\begin{equation}
\small
NP_i = \left \{  p \in NP \mid p \in \text{cube}_i \right \}.
\end{equation}

We represent each point $p \in NP_i$ using a five-dimensional feature vector $\boldsymbol{x}(p)$, which consists of raw signal metrics:
\begin{equation}
\small
\boldsymbol{x}(p) = [snr, \; scr, \; noise, \; sp, \; rp]^{\top}, \quad \forall \, p \in NP_i.
\end{equation}

To mitigate the effects of differing scales and units across signal metrics, we standardize each feature using z-score normalization~\cite{jain2005score}. For cube $i$, we calculate the mean and standard deviation across all noise points:
\begin{equation}
    \small \boldsymbol{\mu_z^i} = [\mu_{snr}^i, \; \mu_{scr}^i, \; \mu_{noise}^i, \; \mu_{sp}^i, \; \mu_{rp}^i]^{\top},
\end{equation}
\begin{equation}
    \small \boldsymbol{\sigma_z^i} = [\sigma_{snr}^i, \; \sigma_{scr}^i, \; \sigma_{noise}^i, \; \sigma_{sp}^i, \; \sigma_{rp}^i]^{\top}.
\end{equation}

The standardized vector $\boldsymbol{z}(p)$ for each point $p \in NP_i$ is then computed as: ${\boldsymbol{z}(p) = \frac{\boldsymbol{x}(p) - \boldsymbol{\mu_z^i}}{\boldsymbol{\sigma_z^i}}}$.

Next, we estimate the parameters of the noise distribution in cube $i$ using Maximum Likelihood Estimation (MLE). For a multivariate Gaussian distribution, the parameters consist of the mean vector ($\hat{\boldsymbol{\mu}}_{i}$) and the covariance matrix ($\hat{\boldsymbol{\Sigma}}_{i}$), which are computed as follows:
\begin{equation}
\small
\hat{\boldsymbol{\mu}}_{i} = \frac{1}{|NP_i|} \sum_{p \in NP_i} \boldsymbol{z}(p),
\end{equation}
\begin{equation}
\small
\hat{\boldsymbol{\Sigma}}_{i} = \frac{1}{|NP_i|} \sum_{p \in NP_i} \left( \boldsymbol{z}(p) - \hat{\boldsymbol{\mu}}_{i} \right) \left( \boldsymbol{z}(p) - \hat{\boldsymbol{\mu}}_{i} \right)^\top.
\end{equation}

Finally, for each cube, we record the z-score standardization statistics and multivariate Gaussian parameters for subsequent noise filtering. The complete parameter set for cube $i$ is given by:
\begin{equation}
\small
\text{Param}_i = \left\{ \boldsymbol{\mu_z^i},\; \boldsymbol{\sigma_z^i},\; \hat{\boldsymbol{\mu}}_{i},\; \hat{\boldsymbol{\Sigma}}_{i} \right\}.
\end{equation}

\subsubsection{Similarity-based Noise Filtering.}

After modeling the noise distribution, we evaluate each point in the current frame for its similarity to the corresponding noise distribution using Mahalanobis distance~\cite{de2000mahalanobis}. This distance metric considers the correlations among features, making it well-suited for multivariate data. Points with high similarity to the noise distribution (i.e., small Mahalanobis distances) are identified as noise and subsequently removed.

Specifically, for each point $p$ in the current frame, we locate its corresponding cube $i$, standardize its feature vector, and compute the Mahalanobis distance as:
\begin{equation}
\small
D(p; \text{Param}_i) = \sqrt{ \left( \boldsymbol{z}(p) - \hat{\boldsymbol{\mu}}_i \right)^\top \hat{\boldsymbol{\Sigma}}_i^{-1} \left( \boldsymbol{z}(p) - \hat{\boldsymbol{\mu}}_i \right) },
\end{equation}
where $\boldsymbol{z}(p)$ is the z-score-normalized feature vector of point $p$, and $\hat{\boldsymbol{\mu}}_{i}$ and $\hat{\boldsymbol{\Sigma}}_{i}$ denote the estimated mean vector and covariance matrix of the noise distribution in cube $i$, respectively.

We then classify a point as noise and remove it if its Mahalanobis distance falls below a similarity threshold $\tau_{sim}$. This threshold is determined adaptively based on the distribution of Mahalanobis distances computed from the historical noise samples $NP$:
\begin{equation}
\small
\tau_{sim} = \operatorname{Percentile}_k \left( \left\{ D(p; \text{Param}_i) \mid p \in  NP \right\} \right),
\end{equation}
where $\operatorname{Percentile}_k(\cdot)$ returns the $k$-th percentile of the computed distances. Empirically, we set $k=80$, which means 80\% of historical noise points in $NP$ have a Mahalanobis distance below the threshold. This allows us to filter most of the noise points while mitigating the impact of outliers.

\begin{figure}
    \centering
    \setlength{\abovecaptionskip}{1mm}
    \includegraphics[width=0.99\linewidth]{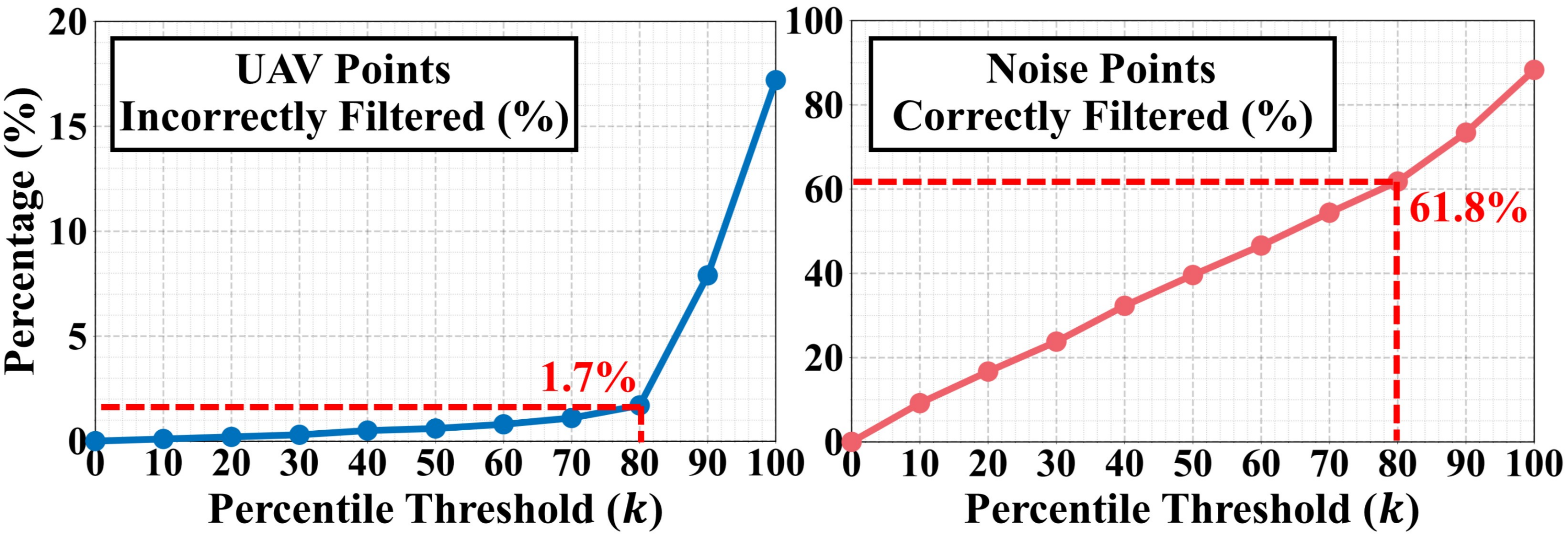}
    \caption{Percentage of filtered UAV points and noise points at different $k$ values. \textmd{At $k=80$, we achieve a good balance, filtering 61.8\% of noise points while filtering only 1.7\% of UAV points.}}
    \label{fig: filtering_k}
    \vspace{-3mm}
\end{figure}

Finally, to evaluate the impact of the threshold parameter $k$, we plot in Figure~\ref{fig: filtering_k} the percentages of filtered UAV points and noise points at different $k$ values. As $k$ increases, more noise points are removed, but the risk of incorrectly filtering UAV points also rises. At $k=80$, we achieve a good balance---removing 61.8\% of noise points while only filtering 1.7\% of UAV points---making it a suitable choice for our system.

\subsubsection{Automatic Model Parameter Updating.}
\label{sec: parameter_update}
Over time, the signal reflection patterns in the environment may gradually shift, rendering the previously fitted Gaussian parameters inaccurate. 
To address this issue, after model initialization, the noise model is automatically updated using recent point cloud data with detected UAV points excluded. 
This design improves model adaptability, enabling continuous operation in real-world environments.
Specifically, we perform parameter updates at fixed intervals of $T$. At each update, we retrospectively collect the most recent $N$ frames of point cloud data, denoted as $P_{\text{upd}}$. 
Since $P_{\text{upd}}$ may contain true UAV points that could contaminate the noise model, we leverage the system’s final detection results to identify and exclude these points.
The remaining subset, denoted as $NP_{\text{upd}}$, contains only noise points and is used to refit the multivariate Gaussian models and recompute the similarity threshold.
In practice, we find that updating the model once every two days is sufficient, and each update requires fewer than 700 frames (approximately 7.5 minutes) of data. 
The choice of optimal values for $T$ and $N$ will be discussed in Section~\ref{sec: noise_point_filtering_performance}.

\vspace{4pt} \noindent \textit{\textbf{Key Result:} After this module, the average number of noise points per frame reduces from \textbf{168.05} to \textbf{64.15}, achieving a \textbf{61.8\%} reduction, as detailed in Section~\ref{sec: noise_point_filtering_performance}.}

\subsection{False Object Filtering Using Spatial- and Velocity-based Confidences}
\label{sec: fake_object_filtering}
After leveraging intrinsic point-level signal metrics, we further analyze the cross-frame motion patterns to eliminate false objects.
These are typically caused by multi-path effects and sidelobe interference from moving vehicles, and often appear as mid-air points at altitudes of 100–300 m.
Our key idea is to perform a consistency check between consecutive frames based on intuitive motion principles. (1) \emph{Spatial consistency:} A true object exhibits continuous spatial motion, so a corresponding object should appear within a local spatial neighborhood in the previous frame. 
(2) \emph{Velocity consistency:} A true object's Doppler velocity measured by the base station should align with the radial velocity computed from the inter-frame spatial displacement.

Building on these consistency principles, we leverage the revealed motion-pattern differences to eliminate false objects.
First, we cluster points into objects.
Then, we compute robust spatial- and velocity-based confidences by aggregating binary consistency outcomes over time using an exponentially decaying weighting strategy inspired by eligibility traces in reinforcement learning~\cite{sutton1998reinforcement}.
Finally, objects with low confidence scores are filtered out.



\subsubsection{Point Clustering.}
Since motion characteristics are exhibited by the object as a whole, we first extend the DBSCAN algorithm~\cite{ester1996density} to cluster points into individual objects more accurately. 
A cluster is formed when a point has sufficient neighboring points within a distance threshold $\tau_d$. 
To avoid the erroneous merging of spatially adjacent objects, we incorporate Doppler velocity into the distance metric. 
The resulting metric is defined as:
\begin{equation}
\small
d = \left\| \left( a_x(x_i - x_j),\, a_y(y_i - y_j),\, a_z(z_i - z_j),\, a_v(v_i - v_j) \right) \right\|_2.
\end{equation}
Here, we empirically set the weights to $(a_x, a_y, a_z, a_v) = (1, 1, 1, 3)$, and the distance threshold to $\tau_d = 10$.
Given the small size of UAVs, which may appear as a single point, we also retain clusters that contain only one point.

\begin{figure}
    \centering
    \setlength{\abovecaptionskip}{1mm}
    \includegraphics[width=0.95\linewidth]{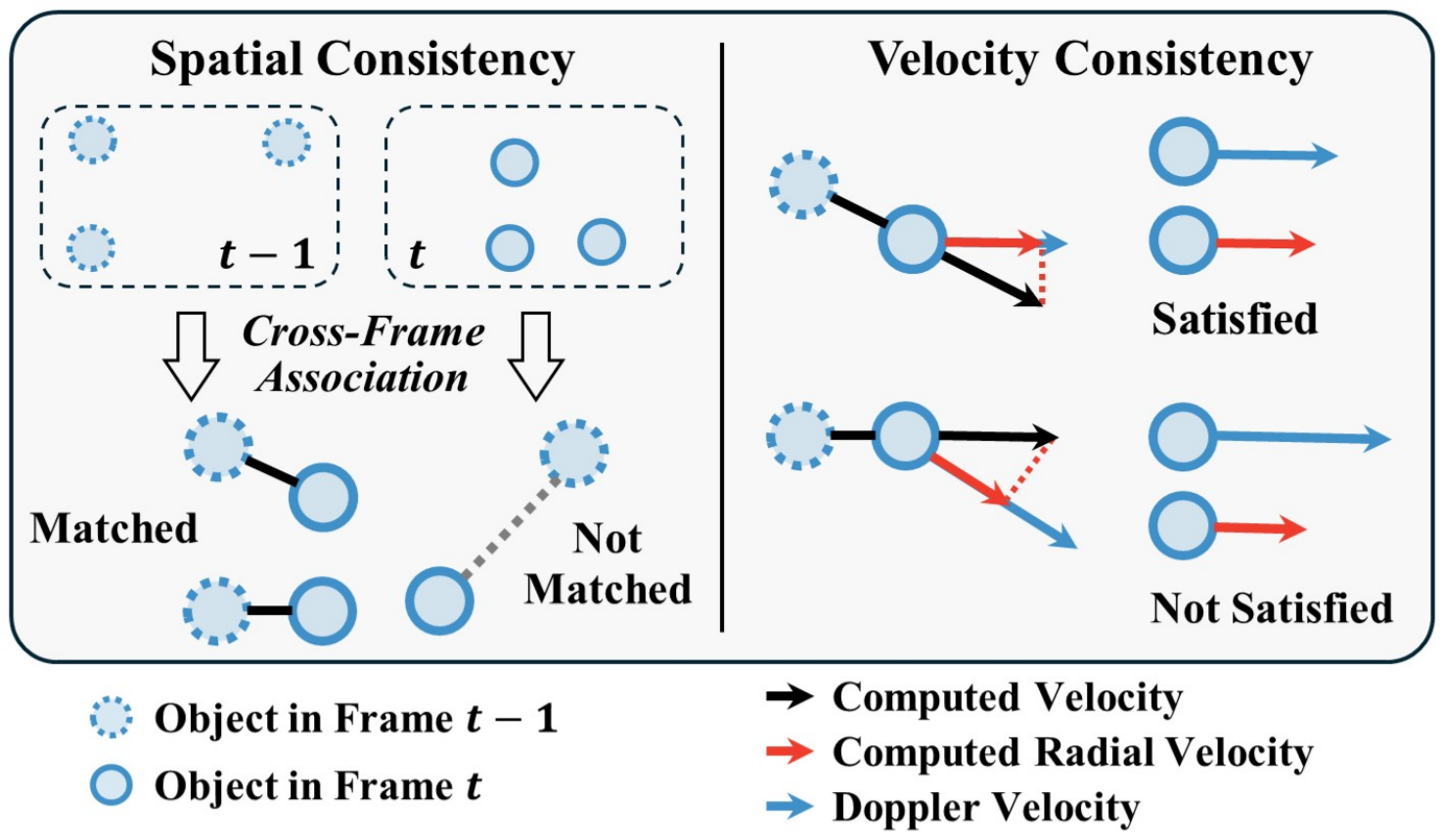}
    \caption{Spatial consistency \textmd{is satisfied when an object is successfully matched across frames.} Velocity consistency \textmd{is satisfied when the radial velocity computed from spatial displacement is consistent with the Doppler velocity.}}
    \label{fig: consistency}
    \vspace{-3mm}
\end{figure}

\subsubsection{Spatial-based Confidence.} 

We design this confidence based on the principle that true targets move continuously through space and exhibit spatial consistency.
As illustrated in Figure~\ref{fig: consistency}, after obtaining the objects from clustering, we employ the Kuhn-Munkres matching algorithm~\cite{kuhn1955hungarian} to establish spatial associations between objects across neighboring frames. 
Two objects are considered a match if their Euclidean distance is less than 15~m. Given the 0.64~s interval between consecutive frames, this threshold is sufficient to cover the inter-frame displacement of commercial UAVs~\cite{Mavic3} operating at typical maximum speeds (below 20~m/s).

For each object $obj$ in the $t$~th frame, we set $I_s(obj, t) = 1$ if a match is identified; otherwise, $I_s(obj, t) = 0$. To quantify how consistently an object is matched over time, we define the \emph{spatial-based confidence} $C_s$ as:
\begin{equation}
\small
C_s\left(obj, t\right) = \gamma_s \cdot \left(C_s(obj', t-1) + I_s(obj, t)\right),
\end{equation}
where $\gamma_s \in [0,1]$ is a decay factor that controls the forgetting rate of historical information (empirically set to 0.9 following~\cite{duan2022pfilter}), and $obj'$ denotes the object in frame $(t-1)$ matched to $obj$.
This recursive formulation updates the confidence of each object over time, giving more weight to recent matches. As a result, the confidence score captures both the frequency and recency of successful associations, providing a robust measure of spatial consistency.

Since true targets move continuously in space and exhibit spatial consistency, a higher confidence score $C_s$ indicates a higher probability that the object is a true target.

\subsubsection{Velocity-based Confidence.}
In addition to spatial consistency, true targets are expected to exhibit velocity consistency: the Doppler velocity measured by the base station should align with the radial velocity inferred from spatial displacement.
Specifically, we first compute the radial velocity of the object relative to the base station based on the spatial association between consecutive frames:
\begin{equation}
\small
v_r = \frac{\mathbf{pos}(obj, t) - \mathbf{pos}(obj', t-1)}{\Delta t} \cdot \frac{\mathbf{pos}(obj, t)}{\| \mathbf{pos}(obj, t) \|},
\end{equation}
where $\mathbf{pos}(obj, t)$ and $\mathbf{pos}(obj', t-1)$ denote the positions of the matched object in frames $t$ and $(t-1)$, respectively.

We then evaluate the consistency between the computed radial velocity $v_r$ and the Doppler velocity $v_d$ measured by the base station, as illustrated in Figure~\ref{fig: consistency}. This consistency is quantified using the following formulation:
\begin{equation}
\small
I_v(obj, t) = \left( |v_r - v_d| < \tau_{c1} \right) \land \left( \frac{\max(v_r, v_d)}{\min(v_r, v_d)} < \tau_{c2} \right) \land \left( v_r \cdot v_d > 0 \right),
\end{equation}
where we empirically set $\tau_{c1} = 2$ and $\tau_{c2} = 2$. If the condition is satisfied, $I_v(obj,t)=1$; otherwise, $I_v(obj,t)=0$. 

Following the same recursive formulation as the spatial confidence, we define the \emph{velocity confidence} $C_v$ as:
\begin{equation}
\small
C_v\left(obj, t\right) = \gamma_v \cdot \left(C_v(obj', t-1) + I_v(obj, t)\right),
\end{equation}
where $\gamma_v$ is also empirically set to 0.9 following~\cite{duan2022pfilter}.

Since the Doppler velocity measured by the base station is expected to be consistent with the radial velocity computed from the spatial displacement of a true target, a higher $C_v$ score indicates a greater likelihood that the object corresponds to a true target.

\subsubsection{Confidence-based Object Discrimination.} 
\begin{figure}
    \centering
    \setlength{\abovecaptionskip}{1mm}
    \includegraphics[width=0.95\linewidth]{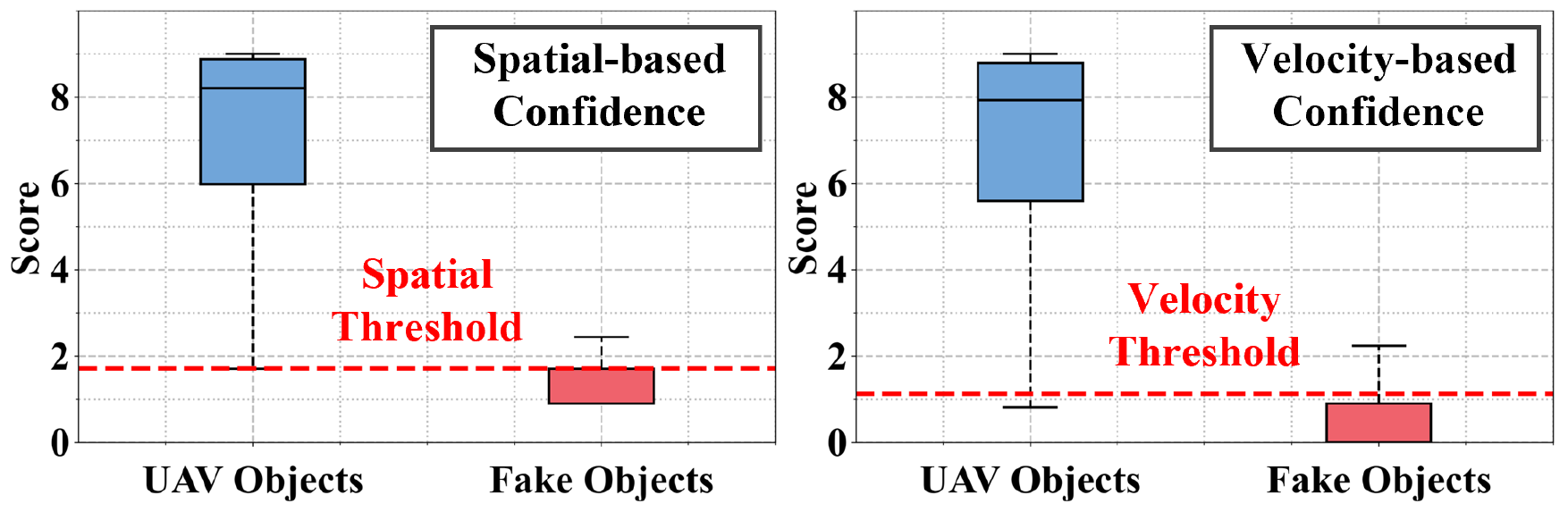}
    \caption{Statistical ranges of spatial- and velocity-based confidences. \textmd{Thresholds are set at the 5th percentile of the UAV confidence scores, with $\tau_{s} = 1.7$ and $\tau_{v} = 1.1$.}}
    \label{fig: confidence_threshold}
    \vspace{-3mm}
\end{figure}

After computing the spatial- and velocity-based confidence scores, we leverage these measures to discriminate between true and false objects. 
Objects that fail to meet the spatial–velocity confidence criteria, specifically those with $C_s < \tau_{s}$ and $C_v < \tau_{v}$, are filtered out.
The thresholds $\tau_{s}$ and $\tau_{v}$ are critical parameters that must be carefully chosen to balance the suppression of false objects with the retention of true targets.
To determine appropriate thresholds, we analyze the confidence score distributions of both UAVs and false objects. As shown in Figure~\ref{fig: confidence_threshold}, both of the proposed confidences provide clear separation between UAVs and false objects. 
Accordingly, we adopt a percentile-based thresholding strategy: to retain more UAVs, we set the thresholds at the 5th percentile of the UAV confidence scores ($\tau_s = 1.7$ and $\tau_v = 1.1$). 
This configuration ensures high recall for UAVs while maintaining effective filtering of false objects.


\vspace{4pt} \noindent \textit{\textbf{Key Result:} After this module, the average number of false objects per frame reduces from \textbf{61.34} to \textbf{2.58}, achieving a \textbf{95.79\%} reduction, as detailed in Section~\ref{sec: fake_object_filtering_performance}.}

\subsection{Object Tracking and False Trajectory Filtering}
\label{sec: fake_trajectory_filtering}

After obtaining the object results, we track their positions over time while simultaneously classifying the generated trajectories.
Notably, the vast majority of false detections have already been removed by the point-level and object-level filtering methods.
We therefore propose a lightweight trajectory classification network that operates in parallel with tracking, continuously classifying and filtering out the few remaining false trajectories in real time.

\subsubsection{IMM-UKF-based Object Tracking.} 
To continuously and accurately track the UAV’s position, we employ the Interacting Multiple Model Unscented Kalman Filter (IMM-UKF)~\cite{nie20233d}, which offers robust tracking under diverse motion patterns. In this framework, the IMM dynamically switches among multiple motion models, while the UKF ensures accurate state estimation in nonlinear systems.
In particular, we define the object state vector to include position, azimuth angle, elevation angle, velocity magnitude, and angular velocity, represented as: $\mathbf{S} = [x,\; y,\; z,\; \phi,\; \theta,\; v,\; \omega_{\phi}]$.

We consider two motion models in our IMM framework. The first is the linear Constant Velocity (CV) model, which assumes that an object moves at a constant speed. The second is the nonlinear Constant Turn Rate and Velocity (CTRV) model, which assumes that an object follows a constant turn rate in the horizontal plane while maintaining a constant velocity in the vertical direction. The tracking algorithm maintains each trajectory and updates it based on the matched object observation ($obj$). 
A trajectory candidate is only confirmed and output after it has been consistently associated with $obj$ across $L$ consecutive frames, where we set $L=6$, a value determined through experiments in Section~\ref{sec:window-size}.



\begin{figure}
    \centering
    \setlength{\abovecaptionskip}{1mm}
    \includegraphics[width=0.95\linewidth]{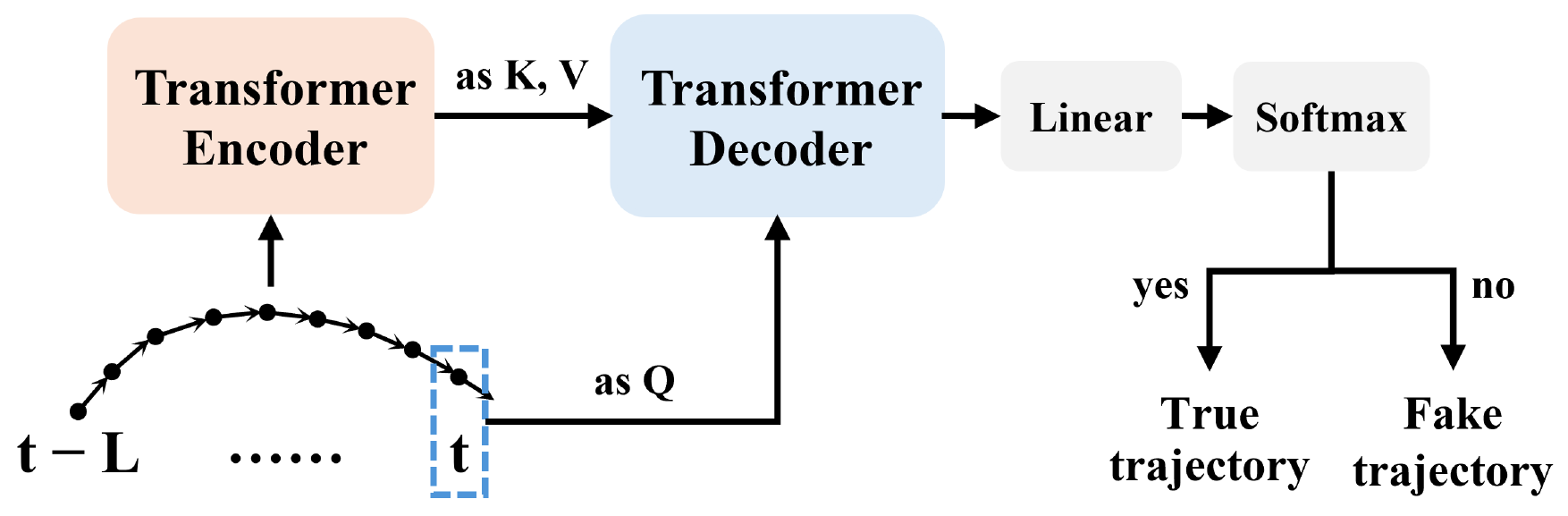}
    \caption{The overview of our Transformer-based trajectory classification network.}
    \label{fig: trajformer}
    \vspace{-3mm}
\end{figure}

\subsubsection{Transformer-based Trajectory Classification.} 
\label{sec: trajformer}

During tracking, false trajectories may still emerge. To address this, we propose TrajFormer, a lightweight Transformer-based trajectory classification network that operates jointly with the tracker to continuously distinguish true trajectories from false ones. Each trajectory is evaluated in real time using the most recent $L$ observations until it is verified as a true target and then output. Each $obj$ comprises spatial motion and signal features: $obj = (x,\; y,\; z,\; v_d,\; snr,\; scr,\; nl,\; sp,\; rp)$.

As shown in Figure~\ref{fig: trajformer}, TrajFormer first employs a Transformer encoder~\cite{vaswani2017attention} to extract temporal features from multi-frame observations ($t-L,\; \ldots,\; t$). 
The current observation ($t$) is then used as the query (Q), while the encoder outputs serve as the key (K) and value (V) in a Transformer decoder~\cite{vaswani2017attention}, which integrates the current frame with its historical context. 
A linear layer followed by a softmax classifier then determines whether the trajectory corresponds to a true target or a false alarm.
The encoder captures long-range temporal dependencies and motion dynamics across consecutive frames, producing a compact representation of trajectory evolution. 
The decoder conditions on the current observation and aligns it with the encoded historical context, thereby highlighting motion consistency or irregularities. 
This design enables the network to learn multi-frame motion patterns and effectively distinguish true trajectories from false ones, including those induced by birds.

\vspace{4pt} \noindent \textit{\textbf{Key Result:} After this module, the average number of false objects per frame decreases from \textbf{2.58} to \textbf{0.04}, and \systemname achieves an \textbf{F1 score of 95.56\%}, as detailed in Section~\ref{sec: fake_trajectory_filtering_performance}.}

\section{Evaluation}
\label{sec: evaluation}

\subsection{Experimental Setup}
\begin{figure}[t]
    \centering
    \setlength{\abovecaptionskip}{1mm}
    \includegraphics[width=0.95\linewidth]{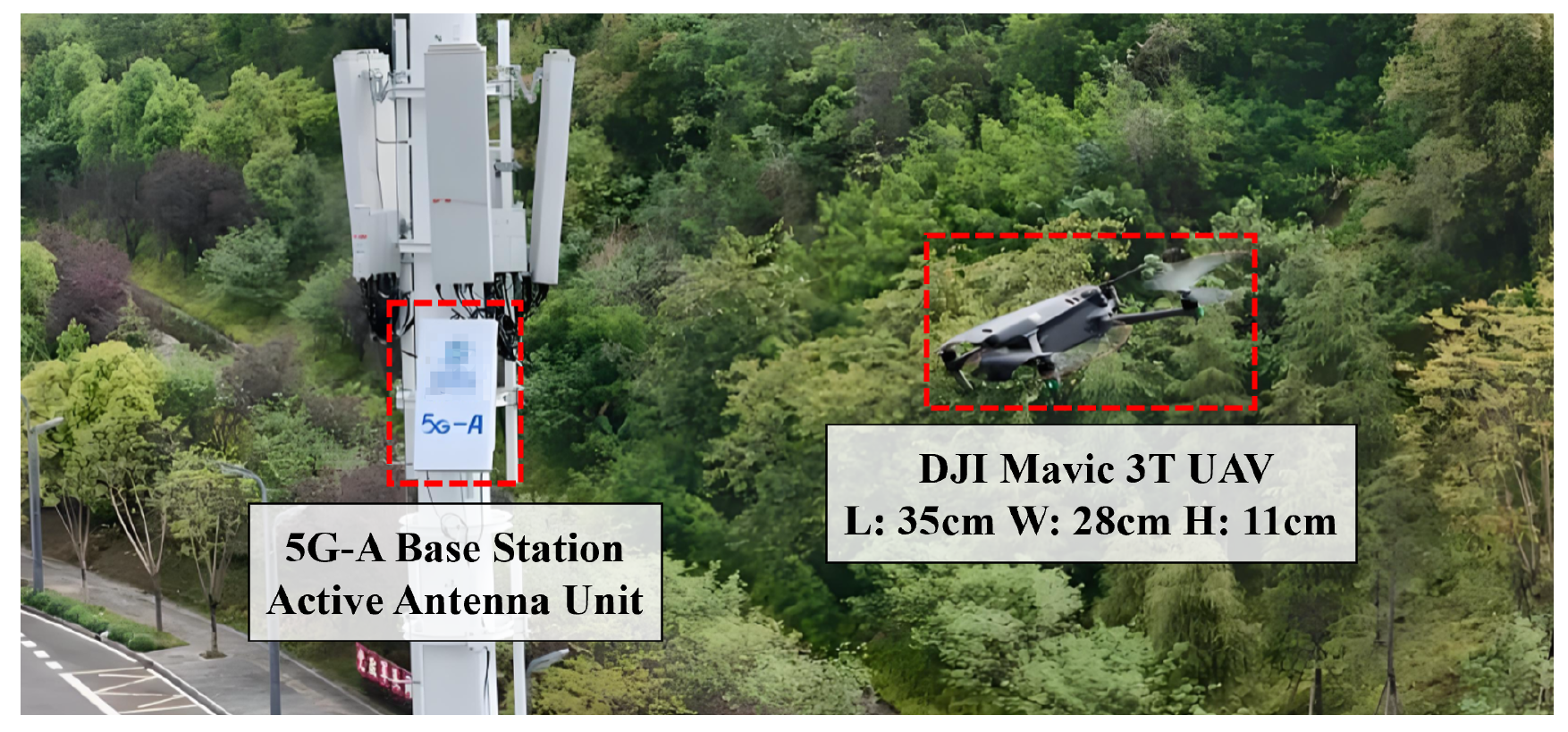}
    \caption{The 5G-A base station Active Antenna Unit (AAU) and the UAV deployed in our field experiments.}
    \label{fig: deployment}
    \vspace{-3mm}
\end{figure}

\noindent \textbf{System Setup.}
We collected data using a Huawei COTS 5G-A base station~\cite{core_network, Huawei2024CloudCore}, operating at 4.9~GHz with a 100~MHz bandwidth and 128 channels, deployed in an urban environment in Shanghai, China.
Figure~\ref{fig: deployment} shows the base station's Active Antenna Unit (AAU) and the DJI Mavic 3T UAV~\cite{Mavic3} used in our field trials.
The AAU was positioned at a height of 23 m, providing a sensing range of up to 1,000 m, with a horizontal field of view (FOV) of $130^\circ$ and a vertical FOV of $45^\circ$.
As shown in Figure~\ref{fig: problem}, the coverage area encompassed residential buildings, factories, overpasses, and a river, providing a complex environment for validating BSense.

\vspace{2pt}
\noindent \textbf{Data Acquisition.}
The UAV was instructed to follow a set of shaped flight paths ($\bigcirc$,\; $\text{S}$,\; $\infty$,\; $\square$,\; $\scalerel*{\diamond}{\square}$,\; $\text{M}$,\; \ding{73}) as well as straight-line paths ($-$) within the sensing area.
For the straight-line flights, the trajectory-boresight angles were set to $-85^\circ$, $-75^\circ$, $\ldots$, $75^\circ$, and $85^\circ$. 
Overall, these paths resulted in 25 distinct trajectories.
The UAV was equipped with GPS, which provided the ground-truth locations for evaluation.
The base station outputs point cloud data at a rate of one frame every 640 milliseconds.
Data were collected continuously over 7 days, resulting in 54 cases with a total duration of approximately 155 minutes (over 14,000 frames). 
Among these, two cases are UAV-free, where no UAV is present. One corresponds to a 10-minute segment at the beginning of the data collection, which is used to initialize the noise fingerprint model (Section~\ref{sec: gaussian_model}), while the other is used to evaluate system performance in target-free scenarios.
In addition, 50 cases are obtained by flying each predefined path type twice, with path lengths ranging from 600~m to 2,300~m.
The remaining two cases consist of a multi-UAV scenario, where two UAVs fly simultaneously, and a cross-site scenario collected from a base station deployed at a different location to evaluate the system’s generalizability.

\vspace{2pt}
\noindent \textbf{Train-Test Split.}
Since the TrajFormer introduced in Section~\ref{sec: trajformer} requires training, we evaluate its performance using 5-fold cross-validation~\cite{wong2019reliable}.
To ensure a fair evaluation and assess generalization ability, the data split is performed at the level of entire flight cases, such that cases with the same trajectory type are never included in both the training and testing folds.
During training, only cases in the training folds are segmented into fixed-size sliding windows to generate positive and negative samples based on ground-truth labels.
In contrast, testing is conducted sequentially on complete trajectories without segmentation.

\vspace{2pt}
\noindent \textbf{Performance Metrics.}
We use the following metrics to evaluate \systemname, focusing on accurate UAV detection and tracking as well as effective filtering of false targets:

\vspace{1pt}
\noindent $\bullet$ \textit{Point-Level Metrics:}
Noise points are first filtered based on noise fingerprint modeling, as described in Section~\ref{sec: noise_point_filtering}. To evaluate this module, we define point-level metrics as follows. A point is classified as a UAV point if its Euclidean distance to the ground truth is within 10~m; otherwise, it is treated as noise.
Let $c_u$ and $c_n$ denote the numbers of UAV and noise points before filtering, and $c_u'$ and $c_n'$ the corresponding numbers after filtering. 
The incorrect filtering rate of UAV points ($F_u$) and the correct filtering rate of noise points ($F_n$) are defined as:
\begin{equation}
\small
F_u = \frac{c_u - c_u'}{c_u}, \quad
F_n = \frac{c_n - c_n'}{c_n}.
\end{equation}

\vspace{1pt}
\noindent $\bullet$ \textit{Object-Level Metrics:}
Then, we obtain object results from the point clustering and false object filtering described in Section~\ref{sec: fake_object_filtering}. For evaluation, we perform minimum-distance binary matching between the detected objects and the ground truth. A detected object is classified as a true positive ($TP$) if the matched distance is within 10~m; otherwise, it is counted as a false positive ($FP$). Ground-truth objects without a match are treated as false negatives ($FN$). 
The UAV recall ($R$), and the average number of false objects per frame ($N_f$) are defined as:
\begin{equation}
\small
R = \frac{TP}{TP + FN}, \quad
N_f = \frac{FP}{\text{total frames}}.
\end{equation}

\vspace{1pt}
\noindent $\bullet$ \textit{Trajectory-Level Metrics:}
Finally, using the object tracking and false trajectory filtering described in Section~\ref{sec: fake_trajectory_filtering}, we evaluate the overall tracking performance, characterized by the F1 score and the mean localization error.
The F1 score is computed as the harmonic mean of precision and recall:
\begin{equation}
\small
P = \frac{TP}{TP + FP}, \quad
R = \frac{TP}{TP + FN}, \quad
F1 = \frac{2PR}{P + R}.
\end{equation}
The mean localization error is defined as:
\begin{equation}
\small
E_{\text{loc}} = \frac{1}{N} \sum_{i=1}^{N} \left\| \hat{\mathbf{pos}}_i - \mathbf{pos}_i \right\|_2,
\end{equation}
which measures the average Euclidean distance between the system-output and ground-truth UAV positions.
Since the primary challenge of this study lies in numerous false detections, we focus particularly on the F1 score.

\begin{figure*}[t]
    \centering
    \setlength{\abovecaptionskip}{1mm}
    \begin{minipage}{0.2601\linewidth}
        \centering
        \includegraphics[width=0.95\linewidth]{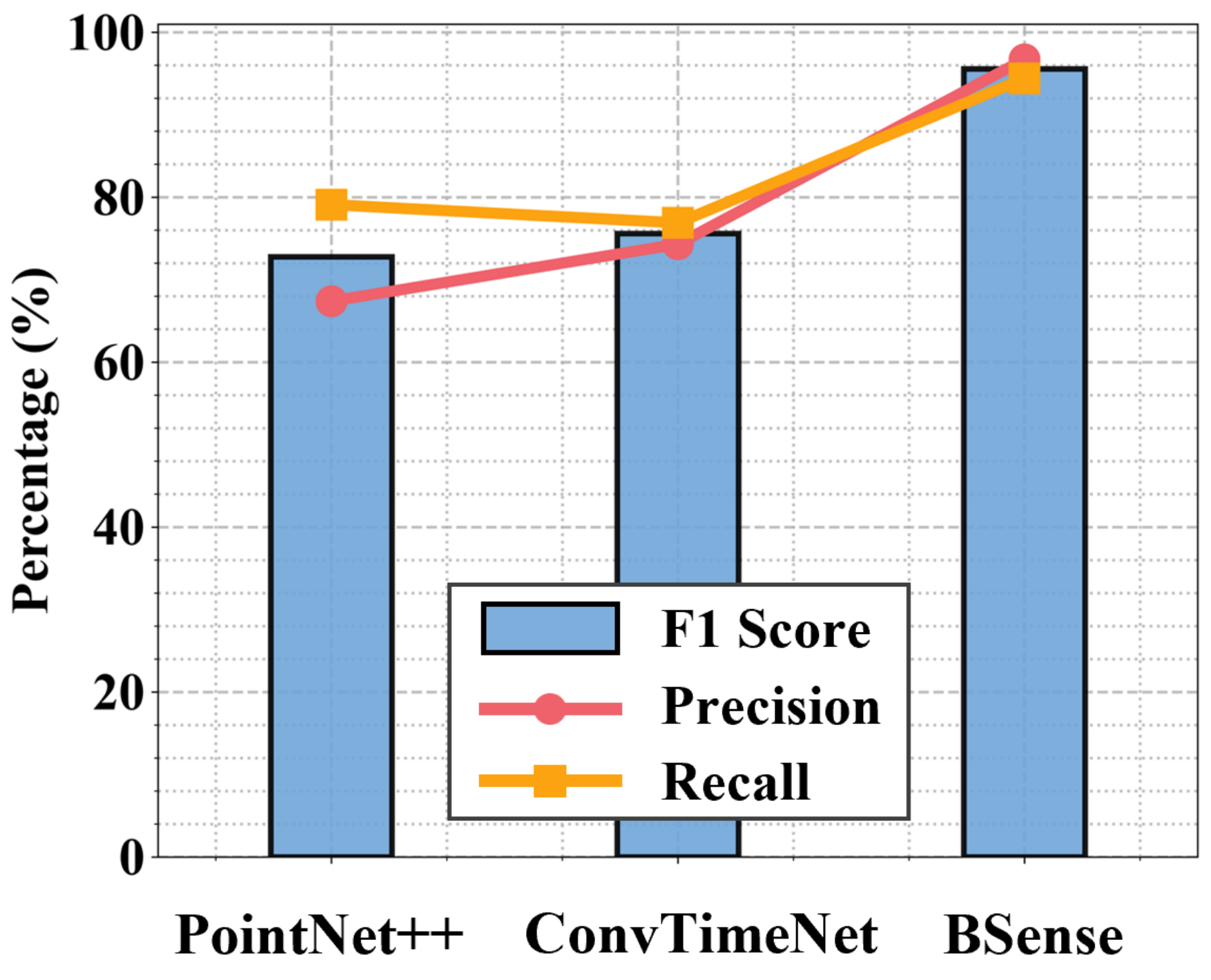}
        \caption{Performance comparison with baselines.}
        \label{fig: result_compare}
    \end{minipage}
    \hspace{10pt}
    \begin{minipage}{0.6899\linewidth}
        \centering
        \includegraphics[width=0.95\linewidth]{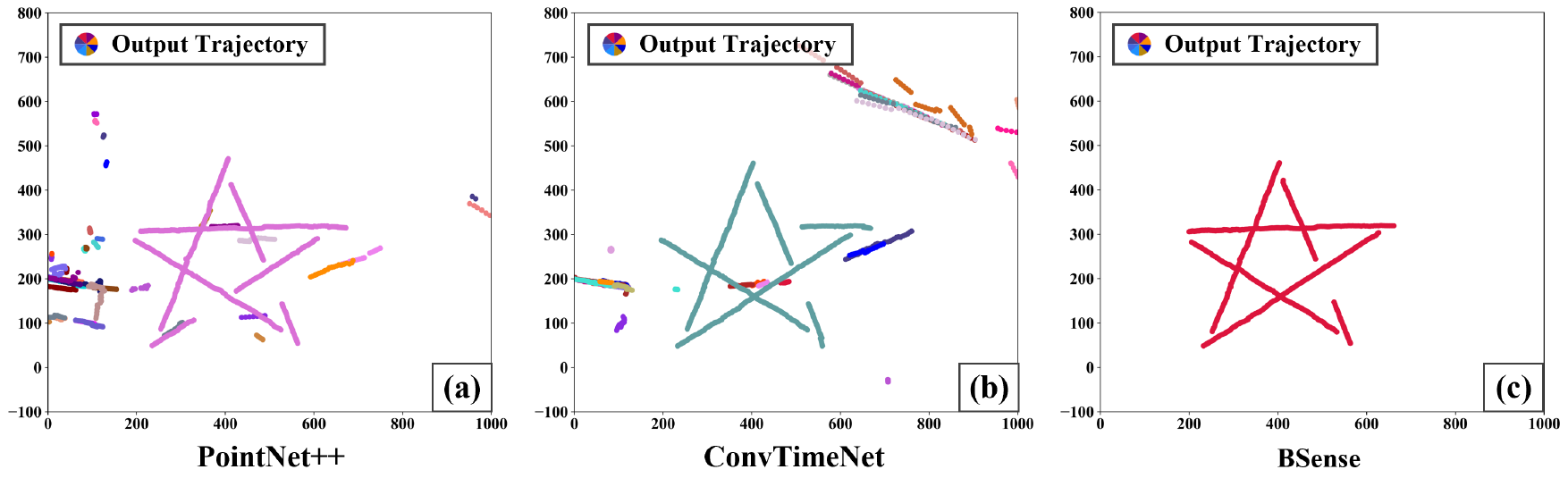}
        \caption{Visual comparison with baselines. \textmd{Compared with (a) and (b), our system (c) filters false trajectories and achieves accurate UAV detection and tracking.}}
        \label{fig: compare_vis}
    \end{minipage}
    \vspace{-3mm}
\end{figure*}

\subsection{Overall Performance}

\noindent \textbf{Comparison with Baselines.}\label{sec:baseline-compare}
Since no existing method targets UAV tracking using point clouds from COTS 5G-A base stations, a direct comparison is not feasible. We therefore compare \systemname with two representative baselines, which filter noise at the point and trajectory levels, respectively.

\begin{figure}[t]
    \centering
    \setlength{\abovecaptionskip}{1mm}
    \includegraphics[width=0.95\linewidth]{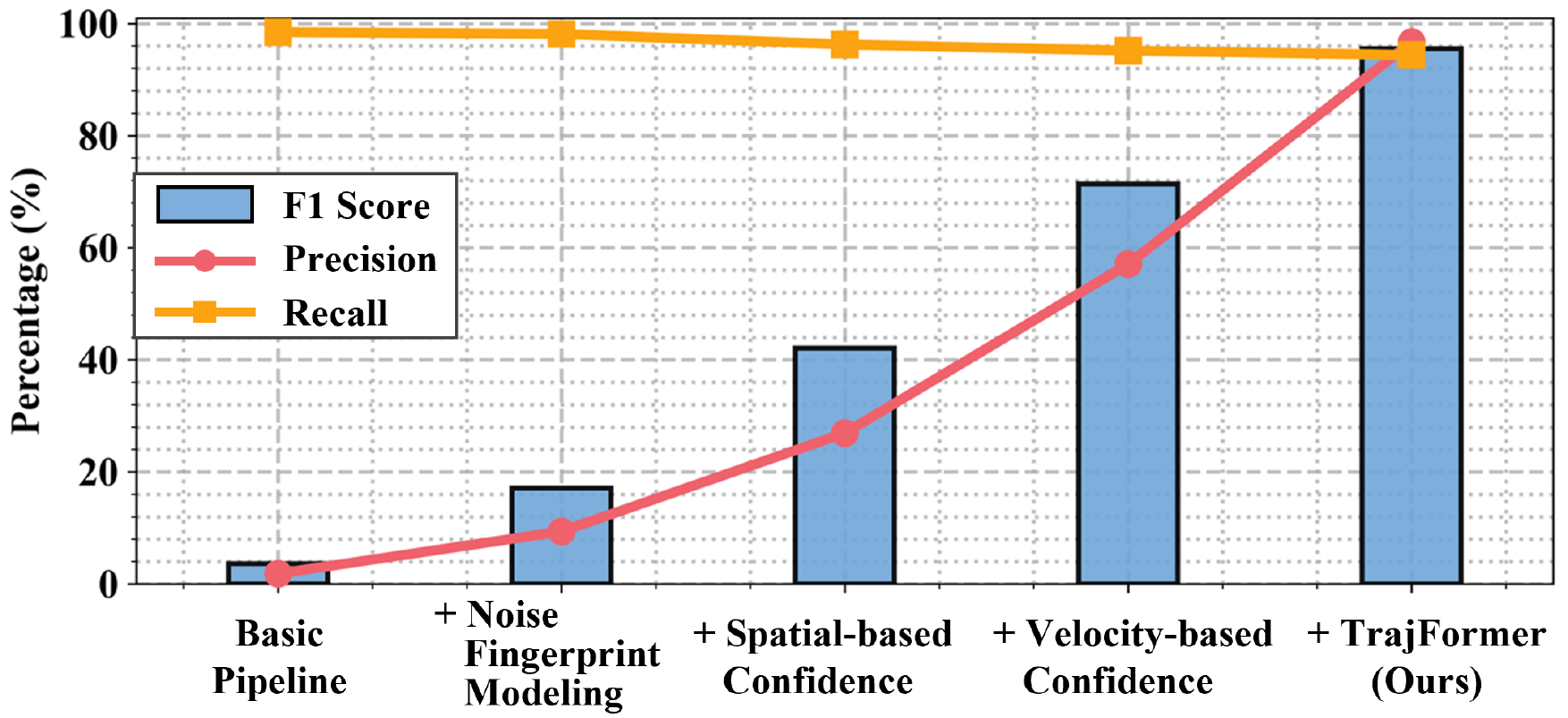}
    \caption{Ablation study. \textmd{The “$+$” indicates stepwise integration of our proposed modules.}}
    \label{fig: result_over_performance}
    \vspace{-3mm}
\end{figure}

\begin{enumerate}[left=0pt]
\vspace{-2pt}
    \item \emph{PointNet++~\cite{qi2017pointnet++}:} A widely used network for classifying radar points~\cite{kopp2023tackling, griebel2021anomaly}. It learns both local and global features from raw point clouds to distinguish and remove false points. The remaining points are then used for clustering and object tracking.
    \item \emph{ConvTimeNet~\cite{cheng2025convtimenet}:} A state-of-the-art time-series classification model. Since trajectories can naturally be treated as time-series data, we first perform point clustering and object tracking, and then apply ConvTimeNet to distinguish true UAV trajectories from false ones.
\end{enumerate}
\vspace{-2pt}
    
As shown in Figure~\ref{fig: result_compare}, BSense achieves the highest F1 score of 95.56\%, which significantly outperforms PointNet++ (72.78\%) and ConvTimeNet (75.59\%).
Figure~\ref{fig: compare_vis} further presents a visual comparison in the \ding{73}-trajectory case, where our system effectively removes false trajectories while accurately preserving and tracking the UAV trajectory.
In contrast, both baselines exhibit noticeable false positives and fragmented trajectories, highlighting the advantage of our approach in achieving accurate and robust UAV tracking.

\vspace{2pt}
\noindent \textbf{Ablation Study.}
We perform an ablation study by incrementally adding our modules to the basic pipeline and evaluating their contributions on the full dataset:

\begin{enumerate}[left=0pt]
\vspace{-2pt}
    \item \emph{Basic Pipeline:} The raw points are directly clustered using DBSCAN~\cite{ester1996density}, and the resulting objects are subsequently tracked using an IMM-UKF~\cite{nie20233d}.
    \item \emph{$+$ Noise Fingerprint Modeling:} Building on \emph{method (1)}, we incorporate noise point filtering based on noise fingerprint modeling, as described in Section~\ref{sec: noise_point_filtering}.
    \item \emph{$+$ Spatial-based Confidence:} Extending \emph{method (2)}, we further introduce spatial-based confidence for filtering false objects, as discussed in Section~\ref{sec: fake_object_filtering}.
    \item \emph{$+$ Velocity-based Confidence:} Extending \emph{method (3)}, we further introduce velocity-based confidence for filtering false objects, as discussed in Section~\ref{sec: fake_object_filtering}.
    \item \emph{$+$ TrajFormer (Ours):} Finally, building on \emph{method (4)}, we employ the proposed TrajFormer network to filter false trajectories, as detailed in Section~\ref{sec: fake_trajectory_filtering}.
\end{enumerate}
\vspace{-2pt}

\begin{figure*}[t]
    \centering
    \setlength{\abovecaptionskip}{1mm}
    \includegraphics[width=0.99\linewidth]{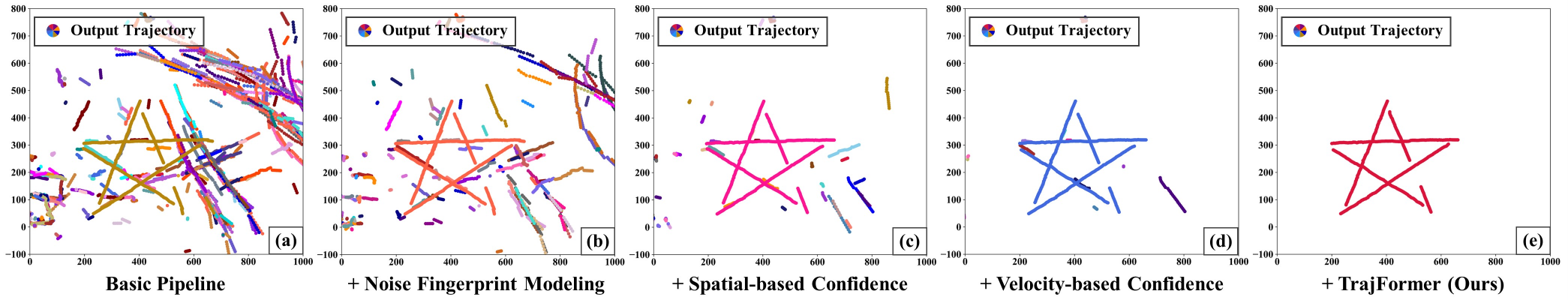}
    \caption{Trajectory visualizations in the ablation study. \textmd{From left to right, “$+$” denotes incremental addition of our proposed modules.}}
    \label{fig: star_vis}
    \vspace{-3mm}
\end{figure*}

\begin{figure*}[t]
    \centering
    \setlength{\abovecaptionskip}{1mm}
    \includegraphics[width=0.99\linewidth]{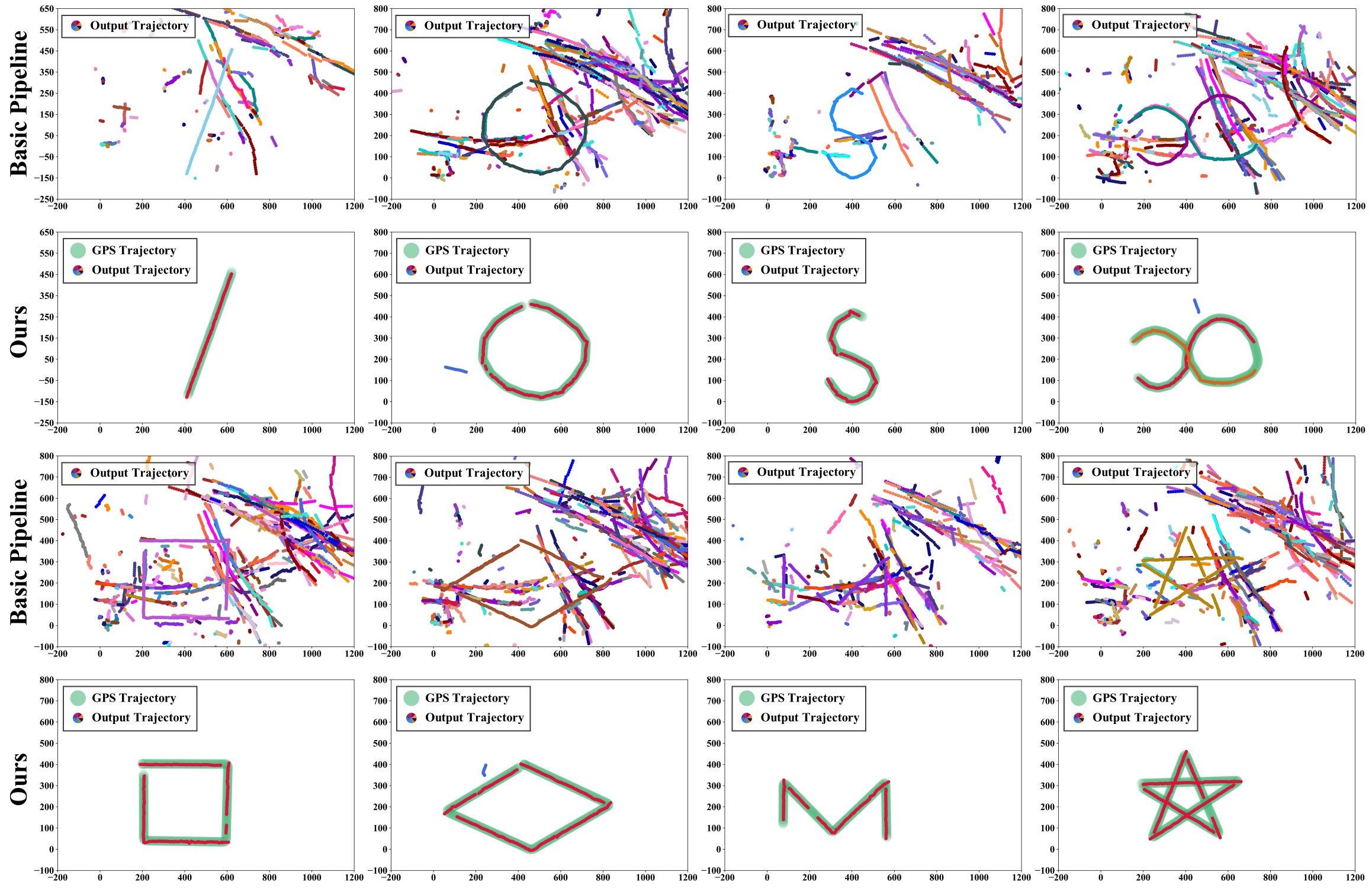}
    \caption{Visual comparison of the basic pipeline and our \systemname across different flight paths. }
    \label{fig: track_vis}
    \vspace{-3mm}
\end{figure*}


Notably, methods (1)–(5) demonstrate the stepwise integration of our proposed modules, with the corresponding results shown in Figure~\ref{fig: result_over_performance}. 
The basic pipeline performs poorly, achieving an F1 score of only 3.65\%, primarily due to numerous noise-induced false positives, which reduce precision to 1.86\%.
Building on this basic pipeline, we progressively incorporate noise fingerprint modeling, spatial-based confidence, velocity-based confidence, and TrajFormer. 
As each module is introduced, precision improves substantially while recall remains high, boosting the F1 score from 3.65\% to 95.56\%.
Overall, this progressive enhancement clearly validates the contribution of each module in our system.

In addition, Figures~\ref{fig: star_vis}(a)–(e) present trajectory visualizations of methods (1)–(5) on the \ding{73}-trajectory case, where different colors denote different trajectories.
As shown, with the incremental integration of our proposed modules, false trajectories are progressively eliminated, while the UAV trajectory is consistently preserved.
This progressive refinement highlights the contribution of each module and further demonstrates the effectiveness of our system in achieving accurate and robust UAV detection and tracking.

\vspace{2pt}
\noindent \textbf{Non-UAV Trajectory Discrimination.}
During data collection, many birds were flying through the sensing area. However, compared to the basic pipeline, BSense filters out 96.66\% of non-UAV trajectories (including birds), demonstrating its ability to distinguish non-UAV flying objects beyond suppressing noise and ghost targets. This is enabled by our trajectory-level model, which captures multi-frame motion patterns and enables accurate discrimination.

\vspace{2pt}
\noindent \textbf{Visualization Results.}
The system’s per-frame outputs are saved and visualized. Figure~\ref{fig: track_vis} presents a comparison between the basic pipeline and our \systemname across different flight paths ($-$,\; $\bigcirc$,\; $\text{S}$,\; $\infty$,\; $\square$,\; $\scalerel*{\diamond}{\square}$,\; $\text{M}$,\; \ding{73}).
The first and third rows show the basic pipeline results, while the second and fourth rows display the corresponding outputs of \systemname.
As illustrated, compared with the basic pipeline, \systemname effectively eliminates false trajectories while accurately detecting and tracking the UAV. 
For the $\infty$, $\text{M}$, and \ding{73} trajectories, partial gaps appear due to the continuous absence of target raw points during certain periods.
Overall, these results highlight the strong performance of \systemname, which suppresses false detections and produces clean UAV trajectories across flight paths ranging from simple ($-$) to complex (\ding{73}).

\vspace{2pt}
\noindent \textbf{Processing Latency.}
In addition to detection and tracking accuracy, low processing latency is critical for real-time deployment. 
We evaluate this by implementing our system on an Intel Core i5-11500 CPU~\cite{i5-11500} and measuring the per-frame latency of each stage.
The results show that the three sequential stages—noise point filtering, false object filtering, and false trajectory filtering—incur average latencies of 6.91 ms, 15.12 ms, and 16.34 ms per frame, respectively. 
Overall, the end-to-end processing latency is 38.37 ms per frame, corresponding to a throughput of approximately 25 frames per second (FPS).
This is substantially lower than the base station’s raw point cloud generation interval (640 ms per frame), indicating that our system comfortably meets real-time requirements.

\subsection{Performance of Noise Point Filtering}
\label{sec: noise_point_filtering_performance}

\begin{figure}
    \centering
    \setlength{\abovecaptionskip}{1mm}
    \includegraphics[width=0.95\linewidth]{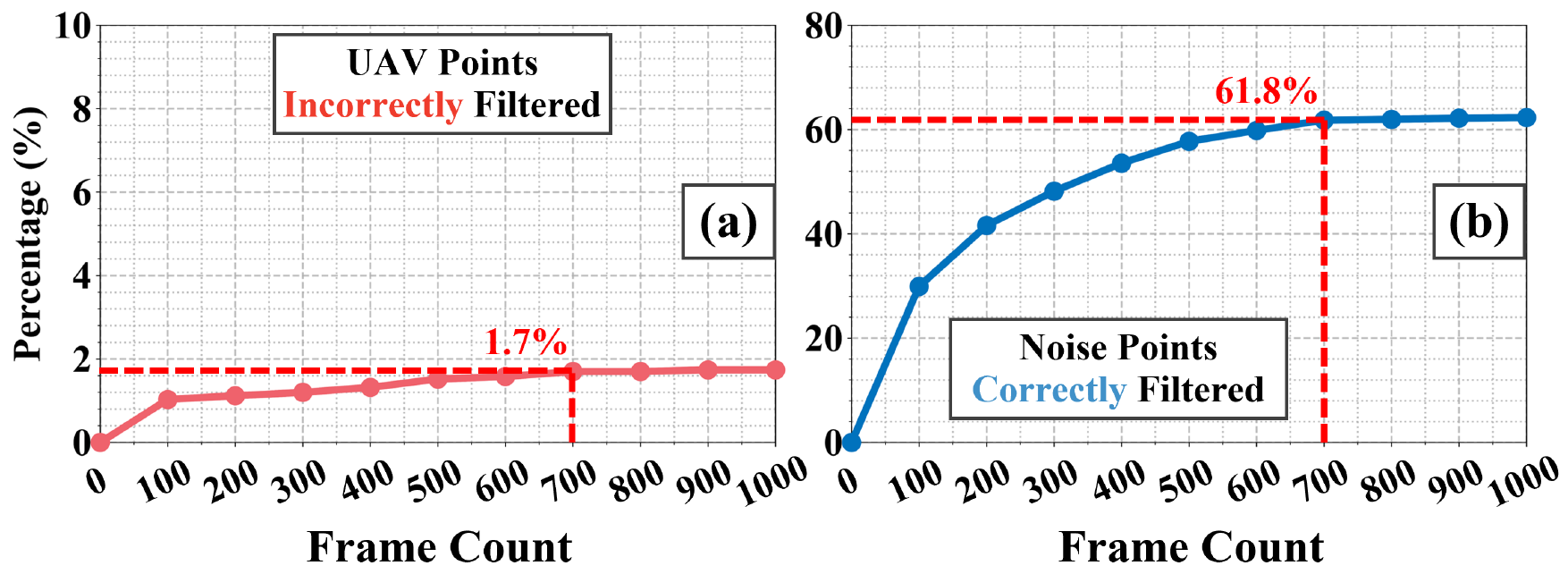}
    \caption{(a) Incorrect filtering rate of UAV points; (b) Correct filtering rate of noise points in different data frame counts.}
    \label{fig: result_pnt_frame}
    \vspace{-3mm}
\end{figure}

\begin{figure}
    \centering
    \setlength{\abovecaptionskip}{1mm}
    \includegraphics[width=0.95\linewidth]{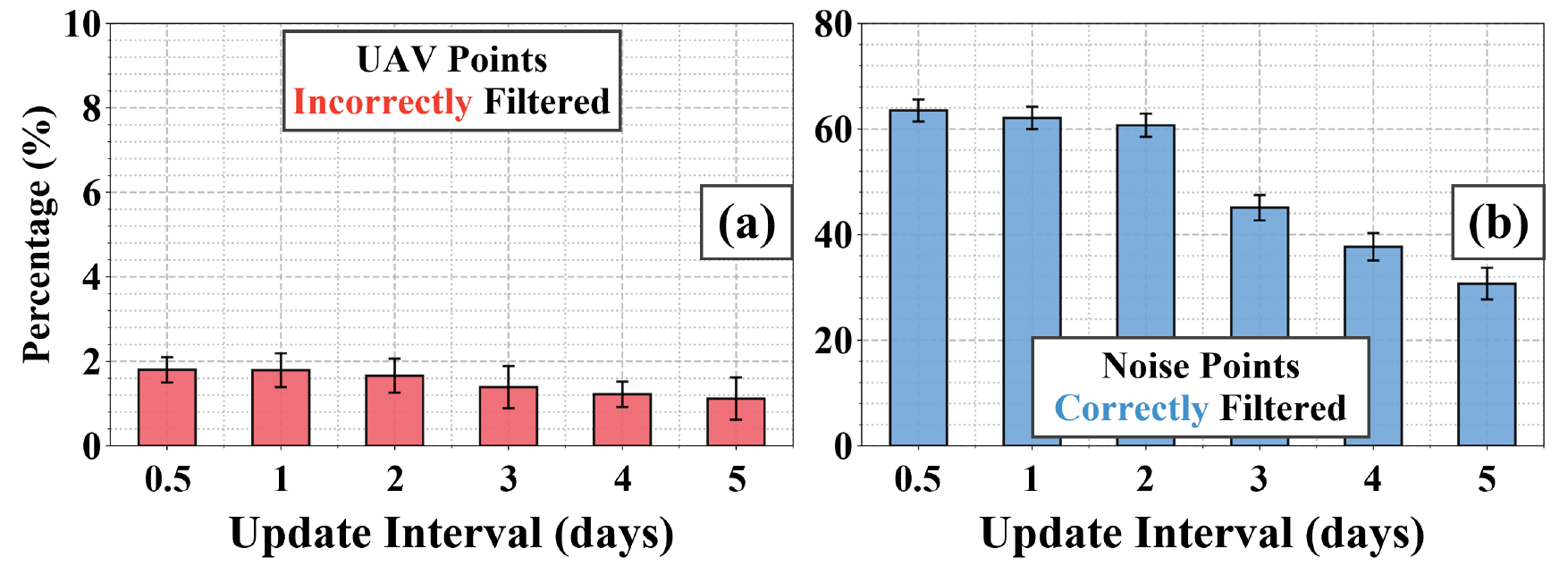}
    \caption{(a) Incorrect filtering rate of UAV points; (b) Correct filtering rate of noise points under different update intervals.}
    \label{fig: result_pnt_update}
    \vspace{-3mm}
\end{figure}

\noindent \textbf{Data Frame Count for Noise Fingerprint Modeling.}
The system adaptively models the noise fingerprint using the most recent $N$ point cloud frames, as described in Section~\ref{sec: noise_point_filtering}. 
Increasing $N$ improves the accuracy of parameter estimation by incorporating more historical observations, but also incurs additional storage overhead.
Figure~\ref{fig: result_pnt_frame} presents the filtering performance under different values of $N$. 
As illustrated in Figure~\ref{fig: result_pnt_frame}(b), the correct filtering rate of noise points increases with larger $N$; however, the marginal gains diminish significantly once $N$ exceeds 700 frames (7.5 minutes).
Meanwhile, Figure~\ref{fig: result_pnt_frame}(a) shows that the incorrect filtering rate of UAV points consistently remains below 2\%, indicating that true UAV points are effectively preserved across different settings.
Considering both filtering effectiveness and storage overhead, we set $N=700$ for noise fingerprint modeling and parameter updates.
Under this configuration, our method reduces the average number of noise points per frame from 168.05 to 64.15, achieving a 61.8\% reduction while incorrectly filtering only 1.7\% of UAV points.

\vspace{2pt}
\noindent \textbf{Model Parameter Update Interval.}
To maintain robust noise point cloud filtering over time, the system periodically updates the parameters of the noise fingerprint model at an interval of $T$, as described in Section~\ref{sec: parameter_update}.
Figure~\ref{fig: result_pnt_update} reports the filtering performance under different values of $T$.
As shown in Figure~\ref{fig: result_pnt_update}(b), the correct filtering rate of noise points decreases as $T$ increases, with a noticeable drop once $T$ exceeds 2.
Meanwhile, Figure~\ref{fig: result_pnt_update}(a) shows that the incorrect filtering rate of UAV points consistently remains below 2\%, indicating that the model effectively preserves true UAV points under different update intervals. 
Based on these observations, we set the update interval to $T=2$, which strikes a good balance between update overhead and filtering performance, ensuring both stability and effectiveness.

\begin{figure}
    \centering
    \setlength{\abovecaptionskip}{1mm}
    \includegraphics[width=0.95\linewidth]{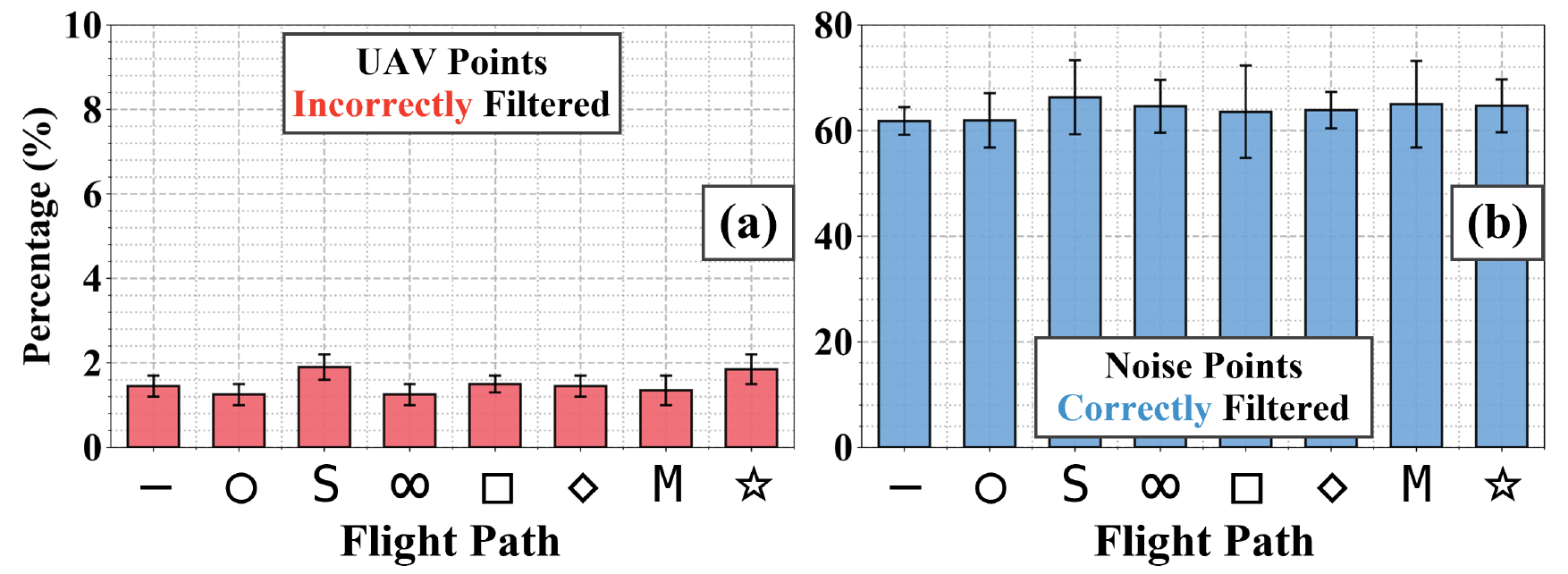}
    \caption{(a) Incorrect filtering rate of UAV points; (b) Correct filtering rate of noise points across different UAV flight paths.}
    \label{fig: result_pnt_shape}
    \vspace{-3mm}
\end{figure}

\vspace{2pt}
\noindent \textbf{Different UAV Flight Paths.}
For UAVs that may appear at any location within the sensing area, it is essential to ensure they are not mistakenly filtered out. Figure~\ref{fig: result_pnt_shape} illustrates the performance of the proposed noise point filtering method across different UAV flight paths ($-$,\; $\bigcirc$,\; $\text{S}$,\; $\infty$,\; $\square$,\; $\scalerel*{\diamond}{\square}$,\; $\text{M}$,\; \ding{73}). As shown in Figure~\ref{fig: result_pnt_shape}(b), the correct filtering rate of noise points consistently exceeds 60\%, while Figure~\ref{fig: result_pnt_shape}(a) shows that the incorrect filtering rate of UAV points remains below 2\% across all flight paths. These results demonstrate the robustness and generalizability of the proposed method, which effectively filters out over 60\% of noise points while preserving more than 98\% of UAV points.

\begin{figure}
    \centering
    \setlength{\abovecaptionskip}{1mm}
    \includegraphics[width=0.95\linewidth]{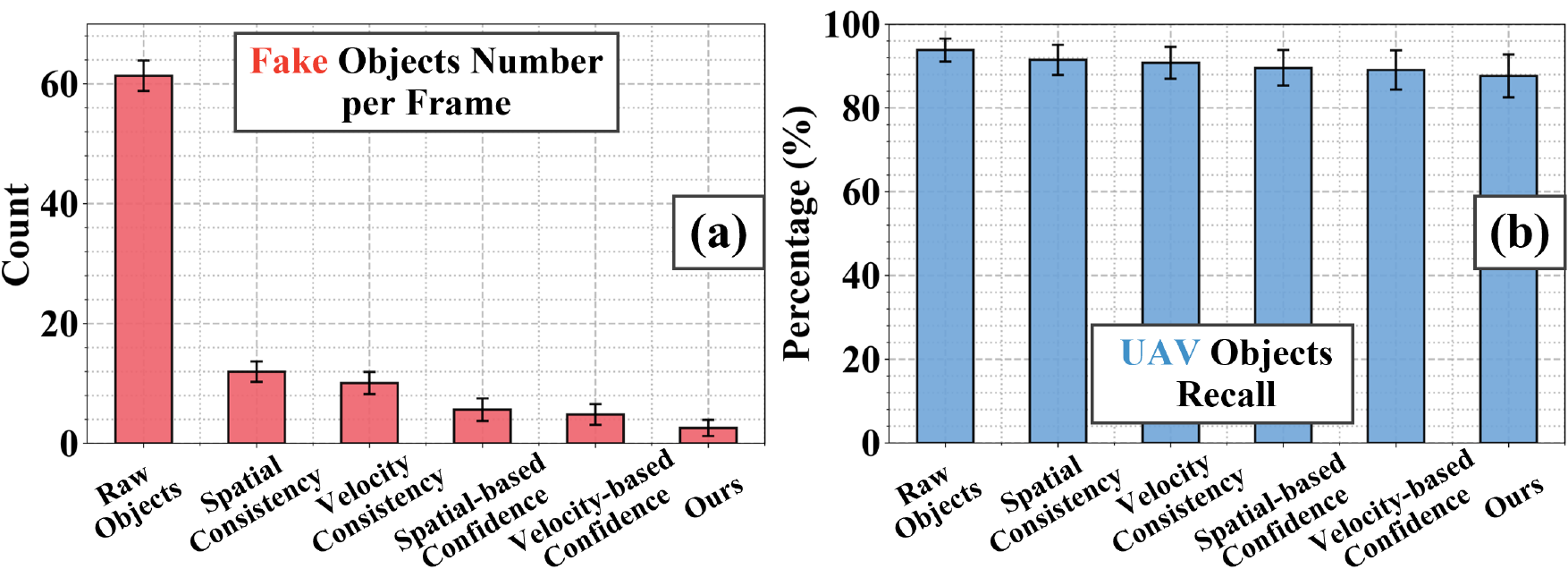}
    \caption{(a) Number of false objects per frame; (b) Recall of UAV objects under varying filtering methods.}
    \label{fig: result_obj_comp}
    \vspace{-3mm}
\end{figure}

\subsection{Performance of False Object Filtering}
\label{sec: fake_object_filtering_performance}

\noindent \textbf{Different Filtering Methods.}
We evaluate the effectiveness of different methods for filtering false objects:
\begin{enumerate}[left=0pt]
\vspace{-2pt}
    \item \emph{Raw Objects:} object results obtained directly from DBSCAN clustering~\cite{ester1996density}.
    \item \emph{Spatial Consistency:} filtering out objects that do not satisfy spatial consistency.
    \item \emph{Velocity Consistency:} filtering out objects that do not satisfy velocity consistency.
    \item \emph{Spatial-based Confidence:} filtering out objects with spatial-based confidence below the threshold.
    \item \emph{Velocity-based Confidence:} filtering out objects with velocity-based confidence below the threshold.
    \item \emph{Ours:} filtering out objects with both spatial- and velocity-based confidences below their thresholds.
\end{enumerate}
\vspace{-2pt}
Figure~\ref{fig: result_obj_comp} compares the performance of these methods. As shown in Figure~\ref{fig: result_obj_comp}(a), applying spatial consistency, velocity consistency, and the derived spatial- and velocity-based confidence significantly reduces the number of false objects.
Meanwhile, Figure~\ref{fig: result_obj_comp}(b) shows that UAV recall remains above 85\% across all methods.
Overall, compared with raw objects, our filtering method using spatial-based and velocity-based confidence reduces the average number of false objects from 61.34 to 2.58 per frame, achieving a 95.79\% reduction while maintaining an 87.64\% UAV recall.
Moreover, UAV recall can be further improved through subsequent object tracking.

\begin{figure}[t]
    \centering
    \setlength{\abovecaptionskip}{1mm}
    \includegraphics[width=0.95\linewidth]{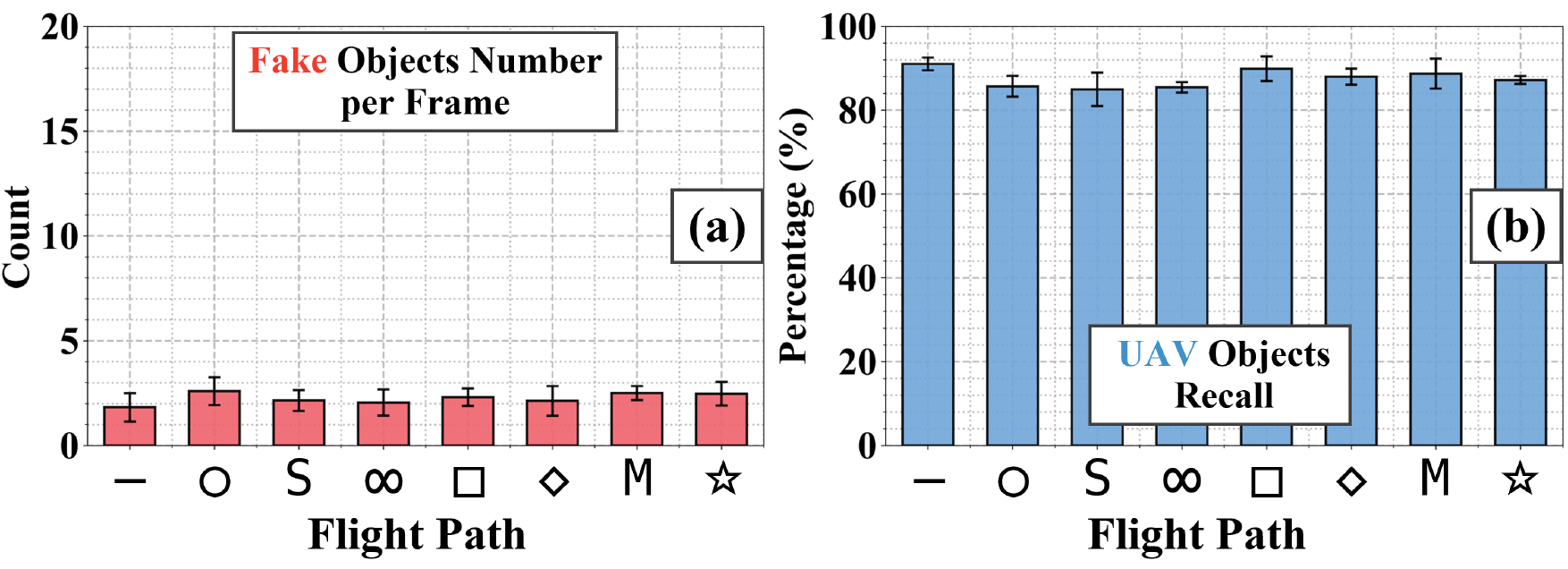}
    \caption{(a) Number of false objects per frame; (b) Recall of UAV objects across different UAV flight paths.}
    \label{fig: result_obj_shape}
    \vspace{-3mm}
\end{figure}

\vspace{2pt}
\noindent \textbf{Different UAV Flight Paths.}
For UAVs that may appear anywhere within the sensing area, it is critical to ensure they are not incorrectly discarded. Figure~\ref{fig: result_obj_shape} illustrates the performance of the proposed false object filtering method across different UAV flight paths ($-$,\; $\bigcirc$,\; $\text{S}$,\; $\infty$,\; $\square$,\; $\scalerel*{\diamond}{\square}$,\; $\text{M}$,\; \ding{73}). As shown in Figure~\ref{fig: result_obj_shape}(a), the number of false objects per frame consistently remains below 3, while Figure~\ref{fig: result_obj_shape}(b) indicates that the recall of UAV objects consistently exceeds 85\% across all flight paths. These results highlight the robustness and adaptability of the proposed method, demonstrating reliable performance under diverse flight scenarios.

\begin{figure}[t]
    \centering
    \setlength{\abovecaptionskip}{1mm}
    \includegraphics[width=0.95\linewidth]{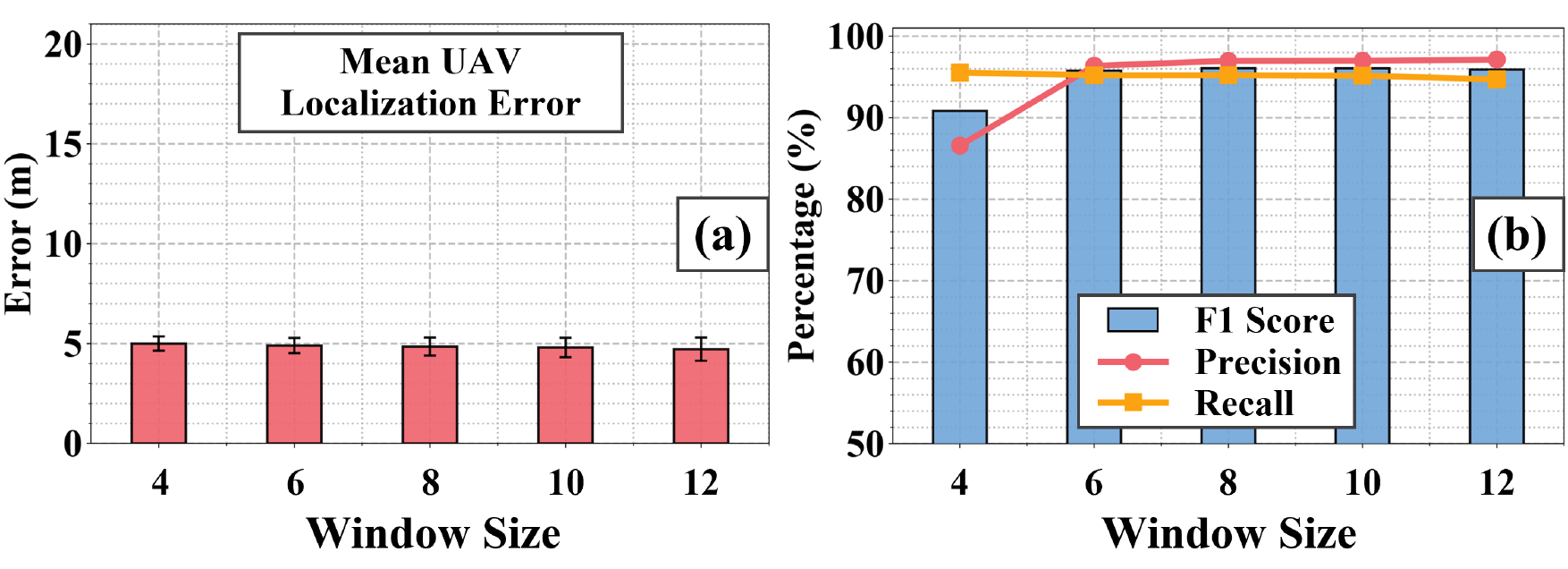}
    \caption{UAV tracking performance under varying window sizes: (a) mean localization error; (b) F1 score.}
    \label{fig: result_trk_len}
    \vspace{-3mm}
\end{figure}

\begin{figure}[t]
    \centering
    \setlength{\abovecaptionskip}{1mm}
    \includegraphics[width=0.95\linewidth]{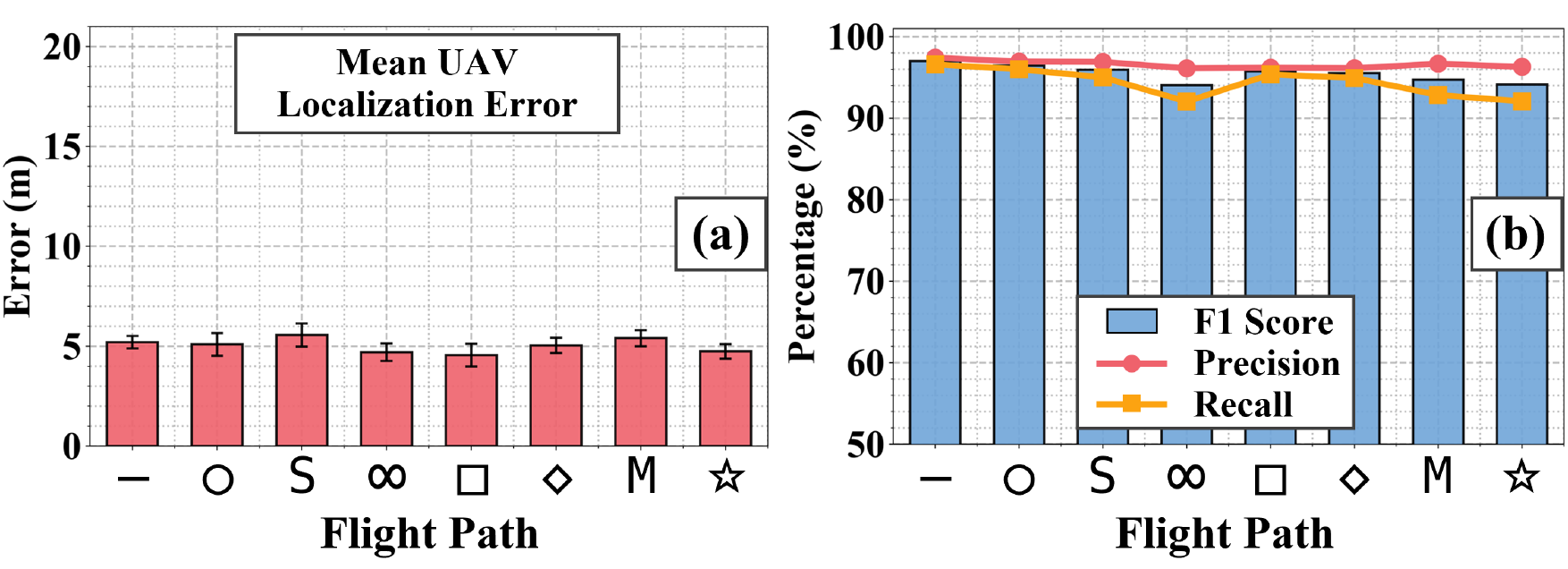}
    \caption{UAV tracking performance across different flight paths: (a) mean localization error; (b) F1 score.}
    \label{fig: result_trk_shape}
    \vspace{-3mm}
\end{figure}

\begin{figure}[t]
    \centering
    \setlength{\abovecaptionskip}{1mm}
    \includegraphics[width=0.95\linewidth]{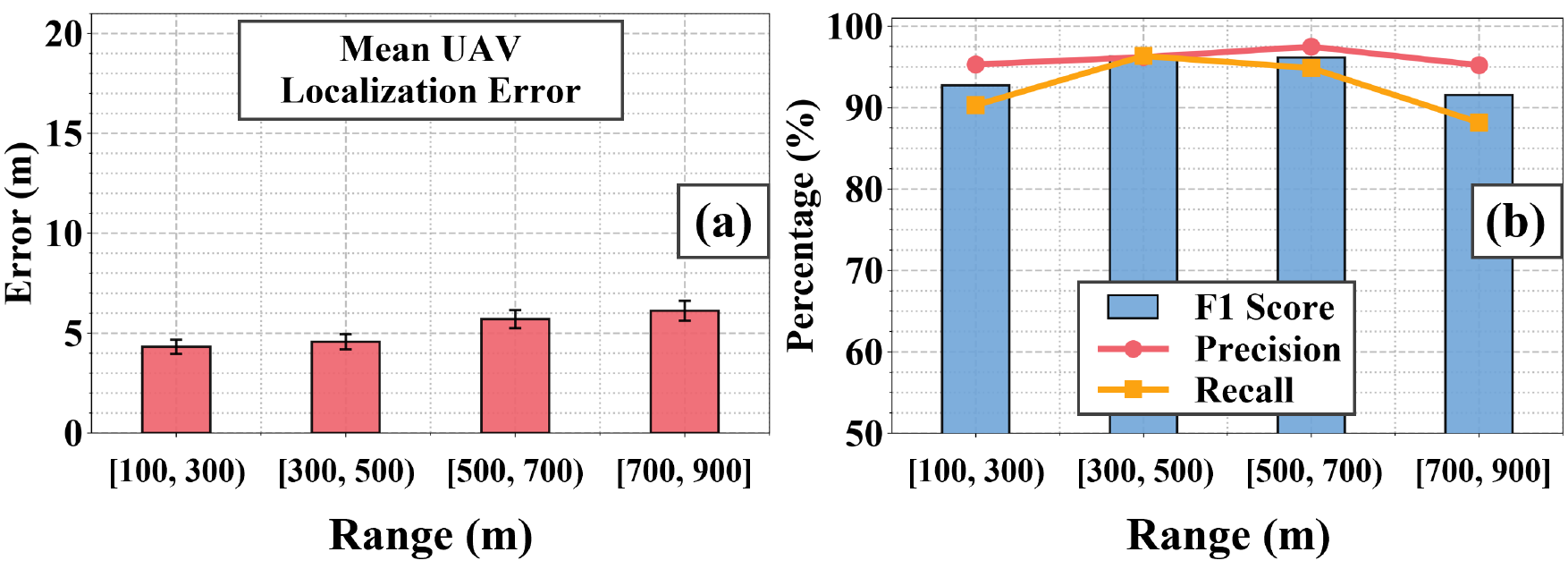}
    \caption{UAV tracking performance across different ranges: (a) mean localization error; (b) F1 score.}
    \label{fig: result_trk_range}
    \vspace{-4mm}
\end{figure}

\subsection{Performance of UAV Tracking}
\label{sec: fake_trajectory_filtering_performance}

\noindent \textbf{Trajectory Classification Window Size.}
\label{sec:window-size}
The system determines whether a trajectory is a true target or a false one using TrajFormer, which processes the most recent $L$ observations, as described in Section~\ref{sec: trajformer}. A larger window size $L$ enables TrajFormer to capture richer temporal information, thereby improving trajectory classification performance. However, increasing $L$ also introduces output latency, as the system must accumulate more observations before making a decision. 
Figure~\ref{fig: result_trk_len} presents the tracking performance across different window sizes. Figure~\ref{fig: result_trk_len}(b) shows that when $L$ exceeds 6, the improvements in F1 score become marginal, while Figure~\ref{fig: result_trk_len}(a) demonstrates that the mean UAV localization error consistently remains below 6~m, satisfying the requirement for kilometer-scale sensing.
Therefore, we set $L=6$ in our experiments.
Under this configuration, the system achieves strong detection performance with an F1 score of 95.56\% (96.73\% precision, 94.41\% recall) and maintains a mean localization error of 4.9~m.

\vspace{2pt}
\noindent \textbf{Different UAV Flight Paths.}
For flight paths ranging from simple ($-$) to complex (\ding{73}), obtaining clean and reliable UAV trajectories is crucial. Figure~\ref{fig: result_trk_shape} presents tracking performance across different flight paths ($-$,\; $\bigcirc$,\; $\text{S}$,\; $\infty$,\; $\square$,\; $\scalerel*{\diamond}{\square}$,\; $\text{M}$,\; \ding{73}). Precision consistently exceeds 96\% across all paths, highlighting the effectiveness of our system. 
However, recall for the $\infty$, $\text{M}$, and \ding{73} trajectories is comparatively lower, primarily due to the continuous absence of target raw points during certain periods (Figure~\ref{fig: track_vis}). 
The corresponding F1 scores are 94.03\%, 94.72\%, and 94.13\%, respectively, while all other paths achieve scores above 95\%. 
This limitation, arising from sustained target point absence, is further discussed in Section~\ref{sec: discussion} as part of our discussion on system constraints.

\vspace{2pt}
\noindent \textbf{Different Ranges.}
As shown in Figure~\ref{fig: result_trk_range}, we evaluate BSense’s tracking performance across different ranges.
Figure~\ref{fig: result_trk_range}(b) shows that from 100~m to 900~m, BSense consistently achieves an F1 score above 90\%.
The relatively lower recall at close range is due to the UAV tracking initialization phase, while the degradation at long range is caused by the absence of target points.
Figure~\ref{fig: result_trk_range}(a) shows that the mean UAV localization error increases with range, but remains within 6.5~m across all ranges.
These results demonstrate that BSense maintains robust tracking performance over long ranges.

\vspace{2pt}
\noindent \textbf{Target-Free Scenarios.}
We evaluate BSense on a separate 15-minute case (approximately 1,400 frames) without UAV flights. 
This case is distinct from the data used for noise fingerprint modeling and is used to assess false alarms in target-free settings.
BSense achieves a false positive rate of 2\%, rarely producing spurious detections even in the absence of true targets. This result demonstrates effective false alarm suppression in no-UAV scenarios.

\begin{figure}[t]
    \centering
    \setlength{\abovecaptionskip}{1mm}
    \includegraphics[width=0.85\linewidth]{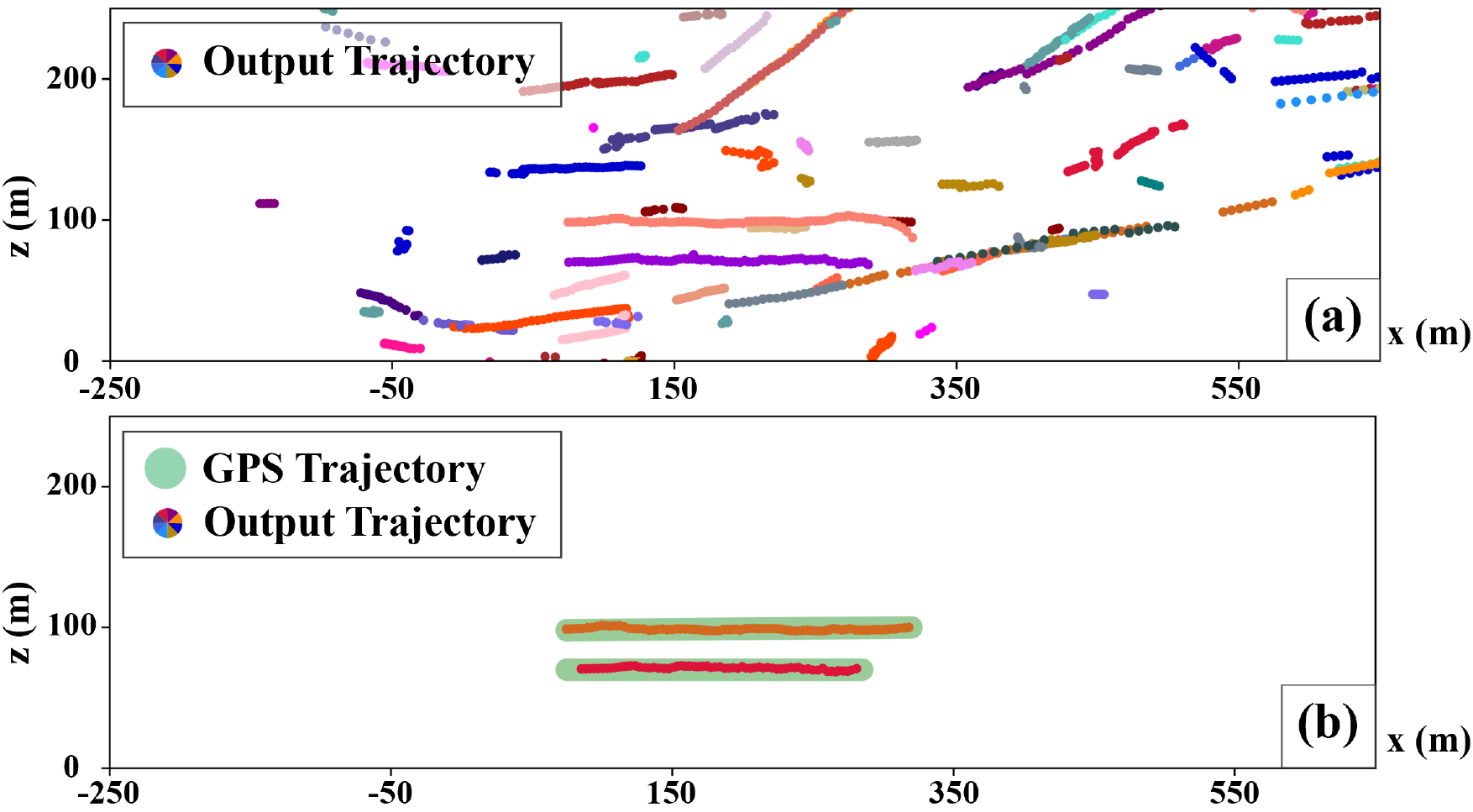}
    \caption{Visual comparison between (a) the basic pipeline and (b) BSense in a multi-UAV scenario.}
    \label{fig: multi_uav}
    \vspace{-3mm}
\end{figure}

\begin{figure}[t]
    \centering
    \setlength{\abovecaptionskip}{1mm}
    \includegraphics[width=0.95\linewidth]{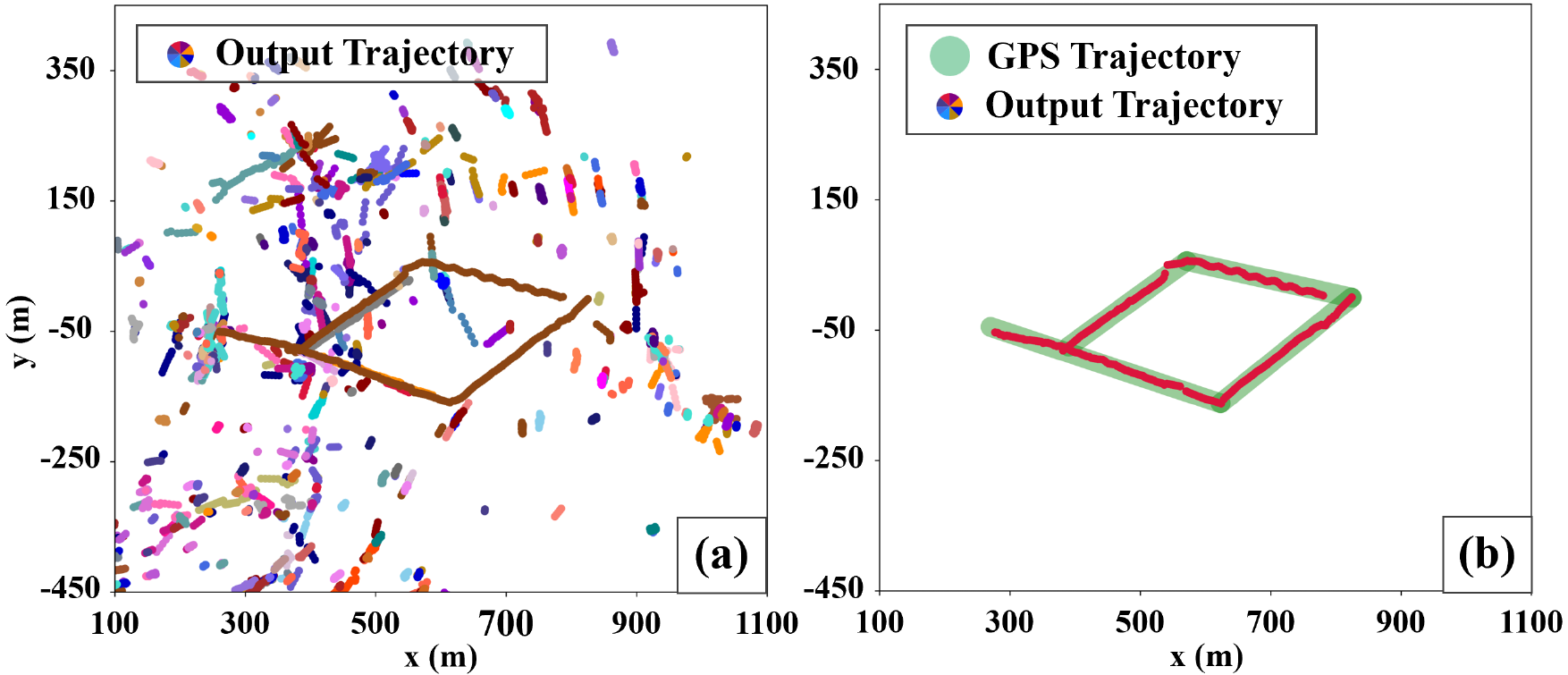}
    \caption{Visual comparison between (a) the basic pipeline and (b) BSense at a different deployment site.}
    \label{fig: cross_site}
    \vspace{-3mm}
\end{figure}

\vspace{2pt}
\noindent \textbf{Multi-UAV Scenario.}
We further evaluate BSense in a multi-UAV scenario in which two UAVs fly in close proximity, thereby introducing additional challenges for target separation.
Figure~\ref{fig: multi_uav} presents the tracking results, comparing the basic pipeline with BSense.
BSense accurately distinguishes and tracks both UAVs, achieving an F1 score of 95.01\% (100\% precision and 90.50\% recall), demonstrating its effectiveness in multi-UAV scenarios.

\vspace{2pt}
\noindent \textbf{Cross-Site Generalization.}
To evaluate the generalizability of BSense, we conduct experiments using data collected from another base station deployed at a different location, where a UAV follows the $\scalerel*{\diamond}{\square}$ flight pattern.
Figure~\ref{fig: cross_site} shows the tracking results, comparing the basic pipeline with BSense.
Consistent with the visual observations, BSense achieves an F1 score of 96.93\% (100\% precision and 93.99\% recall) without any manual retuning, demonstrating robust performance across different deployment scenarios.

\vspace{-4pt}
\section{Discussion}
\label{sec: discussion}

While \systemname achieves strong performance in real-world environments, it still has limitations and open research directions.

\vspace{2pt}
\noindent \textbf{Trajectory Interruption.}
In our experiments, we observe that the UAV trajectories output by the system occasionally exhibit interruptions of 50–100 m, as illustrated in the $\infty$ and \ding{73} examples in Figure~\ref{fig: track_vis}.
These interruptions are not caused by algorithmic misses, but rather by prolonged target absences in the raw point cloud data. 
They typically occur when the UAV is occluded by tall urban buildings or when its flight direction remains tangential to the base station.
In future work, we plan to leverage deep learning to extract motion features from both preceding and succeeding trajectory segments, enabling seamless trajectory completion by generating the missing portions of interrupted trajectories.

\noindent \textbf{Limited Localization Accuracy.}
Our system achieves accurate UAV detection with a mean localization error of 4.9~m.
Although meter-level accuracy at kilometer-scale sensing is nontrivial, further improvement is needed for broader applications.
This work primarily addresses the challenge of filtering massive false detections, while the remaining localization error mostly arises from measurement noise inherent in base-station point clouds; thus, enhancing localization accuracy is beyond our current scope.
Future improvements may include increasing signal bandwidth or expanding the antenna array to boost ranging accuracy and angular resolution, thereby yielding higher-quality point clouds.
\vspace{-4pt}
\section{Related Work}

\noindent \textbf{Sensor-based UAV Sensing.}
Cameras, LiDARs, and radars are widely used for object detection~\cite{chu2025racformer, wang2023bi, meng2024mmplace, meng2025gr, wang2024rdgait, he2024see, he2025ghost}, and recent studies have also explored their use in UAV sensing~\cite{abir2025detection, hammer2018potential, klare2017uav, wang2021deep, zhao2022vision, wu2023precise, seidaliyeva2023advances, wu2024vehicle, huang2023anti, wang2024survey, iizuka2023millisign, zhang2023mmhawkeye}. 
Despite the strong perception performance achieved by these sensor-based approaches, they still suffer from several limitations. Camera-based systems struggle to obtain reliable distance information and therefore typically provide only 2D detection. They are also highly sensitive to lighting conditions~\cite{rezaei2015robust}, making nighttime operation difficult. LiDAR and radar systems, while more robust, tend to be expensive and thus unsuitable for large-scale deployment~\cite{poitevin2017challenges}, limiting their use to specific scenarios such as airports or UAV shows.


\vspace{2pt}
\noindent \textbf{Base Station-based Sensing.}
With the advancement of Integrated Sensing and Communication (ISAC), 5G-A base stations have been explored as sensing platforms in addition to providing communication services~\cite{wei2023integrated, lin2024integrated, wei2024multiple, luo2025isac}.
Recent studies~\cite{zhou2024joint, zhao2024joint, lu2024integrated, elfiatoure2025multiple, li2025joint, yang2025hierarchical,saikia2024hybrid,wang2019assistant,motie2024self,lu2024deep,hu2024joint} have explored the use of 5G-A base stations for vehicle and UAV detection, and environment reconstruction. However, these works primarily focus on ISAC architecture design and signal processing optimization, with evaluations limited to theoretical analysis or idealized simulations. 
To the best of our knowledge, the only real-world deployment reported so far is by Lu et al.~\cite{lu2024intergrated}, which leverages an ISAC base station for maritime vessel detection and tracking. Yet, their experiments were conducted in open environments where large vessels are relatively easy to detect.


\vspace{-4pt}
\section{Conclusion}

We present \systemname, the first practical system that tracks UAVs using commercial 5G-A base stations.
Our system introduces a layered framework that progressively filters noise at the point, object, and trajectory levels.
Through deployment in a complex urban environment, we demonstrate that \systemname reduces hundreds of false detections per frame to nearly zero, while achieving meter-level localization accuracy at kilometer-scale sensing ranges.
We believe that the design, implementation, and evaluation of \systemname mark an important step in advancing ISAC from theoretical exploration to real-world UAV detection and tracking at scale.

\vspace{-4pt}
\begin{acks}
This work was supported by the New Generation Artificial Intelligence-National Science and Technology Major Project (No. 2025ZD0122600), the National Natural Science Foundation of China (No. 62332016), and the Fundamental and Interdisciplinary Disciplines Breakthrough Plan of the Ministry of Education of China (No. JYB2025XDXM113).
\end{acks}

\bibliographystyle{ACM-Reference-Format}
\bibliography{references}

@article{zhou2025unmanned,
  title={Unmanned aerial vehicles based low-altitude economy with lifecycle techno-economic-environmental analysis for sustainable and smart cities},
  author={Zhou, Yuekuan},
  journal={Journal of Cleaner Production},
  pages={145050},
  year={2025},
  publisher={Elsevier}
}

@article{mohamed2020unmanned,
  title={Unmanned aerial vehicles applications in future smart cities},
  author={Mohamed, Nader and Al-Jaroodi, Jameela and Jawhar, Imad and Idries, Ahmed and Mohammed, Farhan},
  journal={Technological forecasting and social change},
  volume={153},
  pages={119293},
  year={2020},
  publisher={Elsevier}
}

@article{zhao2022vision,
  title={Vision-based anti-uav detection and tracking},
  author={Zhao, Jie and Zhang, Jingshu and Li, Dongdong and Wang, Dong},
  journal={IEEE Transactions on Intelligent Transportation Systems},
  volume={23},
  number={12},
  pages={25323--25334},
  year={2022},
  publisher={IEEE}
}

@article{huang2023anti,
  title={Anti-UAV410: A thermal infrared benchmark and customized scheme for tracking drones in the wild},
  author={Huang, Bo and Li, Jianan and Chen, Junjie and Wang, Gang and Zhao, Jian and Xu, Tingfa},
  journal={IEEE Transactions on Pattern Analysis and Machine Intelligence},
  volume={46},
  number={5},
  pages={2852--2865},
  year={2023},
  publisher={IEEE}
}

@inproceedings{abir2025detection,
  title={Detection and Tracking of Drone Swarms using LiDAR},
  author={Abir, Tasnim Azad and Le, Vu and Kuantama, Endrowednes and Gupta, Pranjol Sen and Copley, Austin and Dawes, Judith and Islam, Mohammad and Han, Richard and Nguyen, Phuc},
  booktitle={Proceedings of the 23rd Annual International Conference on Mobile Systems, Applications and Services},
  pages={500--514},
  year={2025}
}

@inproceedings{hammer2018potential,
  title={Potential of lidar sensors for the detection of UAVs},
  author={Hammer, Marcus and Hebel, Marcus and Borgmann, Bj{\"o}rn and Laurenzis, Martin and Arens, Michael},
  booktitle={Laser Radar Technology and Applications XXIII},
  volume={10636},
  pages={39--45},
  year={2018},
  organization={SPIE}
}

@article{wang2021deep,
  title={Deep learning-based UAV detection in pulse-Doppler radar},
  author={Wang, Chenxing and Tian, Jiangmin and Cao, Jiuwen and Wang, Xiaohong},
  journal={IEEE Transactions on Geoscience and Remote Sensing},
  volume={60},
  pages={1--12},
  year={2021},
  publisher={IEEE}
}

@inproceedings{klare2017uav,
  title={UAV detection with MIMO radar},
  author={Klare, Jens and Biallawons, Oliver and Cerutti-Maori, Delphine},
  booktitle={2017 18th International Radar Symposium (IRS)},
  pages={1--8},
  year={2017},
  organization={IEEE}
}

@article{seidaliyeva2023advances,
  title={Advances and challenges in drone detection and classification techniques: A state-of-the-art review},
  author={Seidaliyeva, Ulzhalgas and Ilipbayeva, Lyazzat and Taissariyeva, Kyrmyzy and Smailov, Nurzhigit and Matson, Eric T},
  journal={Sensors},
  volume={24},
  number={1},
  pages={125},
  year={2023},
  publisher={MDPI}
}

@article{li2024farfusion,
  title={FARFusion: A practical roadside radar-camera fusion system for far-range perception},
  author={Li, Yao and Wang, Yingjie and Meng, Chengzhen and Duan, Yifan and Ji, Jianmin and Zhang, Yu and Zhang, Yanyong},
  journal={IEEE Robotics and Automation Letters},
  volume={9},
  number={6},
  pages={5433--5440},
  year={2024},
  publisher={IEEE}
}

@article{wei2023integrated,
  title={Integrated sensing and communication signals toward 5G-A and 6G: A survey},
  author={Wei, Zhiqing and Qu, Hanyang and Wang, Yuan and Yuan, Xin and Wu, Huici and Du, Ying and Han, Kaifeng and Zhang, Ning and Feng, Zhiyong},
  journal={IEEE Internet of Things Journal},
  volume={10},
  number={13},
  pages={11068--11092},
  year={2023},
  publisher={IEEE}
}

@article{luo2025isac,
  title={ISAC--A Survey on Its Layered Architecture, Technologies, Standardizations, Prototypes and Testbeds},
  author={Luo, Xuewen and Lin, Qingfeng and Zhang, Ruoyu and Chen, Hsiao-Hwa and Wang, Xingwei and Huang, Min},
  journal={IEEE Communications Surveys \& Tutorials},
  year={2025},
  publisher={IEEE}
}

@article{zhou2024joint,
  title={Joint Target Detection and Channel Estimation for Distributed Massive MIMO ISAC Systems},
  author={Zhou, Lei and Dai, Jisheng and Xu, Weichao and Chang, Chunqi},
  journal={IEEE Transactions on Cognitive Communications and Networking},
  year={2024},
  publisher={IEEE}
}

@article{zhao2024joint,
  title={Joint Target Localization and Data Detection in Bistatic ISAC Networks},
  author={Zhao, Na and Chang, Qing and Shen, Xiao and Wang, Yunlong and Shen, Yuan},
  journal={IEEE Transactions on Communications},
  year={2024},
  publisher={IEEE}
}

@article{elfiatoure2025multiple,
  title={Multiple-target detection in cell-free massive MIMO-assisted ISAC},
  author={Elfiatoure, Mohamed and Mohammadi, Mohammadali and Ngo, Hien Quoc and Shin, Hyundong and Matthaiou, Michail},
  journal={IEEE Transactions on Wireless Communications},
  year={2025},
  publisher={IEEE}
}

@article{li2025joint,
  title={Joint Target Assignment and Resource Allocation for Multi-Base Station Cooperative ISAC in UAV Detection},
  author={Li, Ruotong and Zhang, Qixun and Ma, Dingyou and Yu, Kan and Huang, Yuzhen},
  journal={IEEE Transactions on Vehicular Technology},
  year={2025},
  publisher={IEEE}
}

@inproceedings{chen2024resilient,
  title={Resilient Massive Access assisted ISAC in Space-Air-Ground Integrated Networks},
  author={Chen, Jiabin and Xie, Wupeng and Wang, Chaowei and Pang, Mingliang and Jiang, Fan and Xu, Lexi},
  booktitle={Proceedings of the 30th Annual International Conference on Mobile Computing and Networking},
  pages={2203--2208},
  year={2024}
}

@article{wei2024multiple,
  title={Multiple reference signals collaborative sensing for integrated sensing and communication system towards 5G-A and 6G},
  author={Wei, Zhiqing and Li, Fengyun and Liu, Haotian and Chen, Xu and Wu, Huici and Han, Kaifeng and Feng, Zhiyong},
  journal={IEEE Transactions on Vehicular Technology},
  volume={73},
  number={10},
  pages={15185--15199},
  year={2024},
  publisher={IEEE}
}

@article{liu2025carrier,
  title={Carrier aggregation enabled MIMO-OFDM integrated sensing and communication},
  author={Liu, Haotian and Wei, Zhiqing and Piao, Jinghui and Wu, Huici and Li, Xingwang and Feng, Zhiyong},
  journal={IEEE Transactions on Wireless Communications},
  year={2025},
  publisher={IEEE}
}

@article{chen2024uitde,
  title={UITDE: A UAV-assisted intelligent true data evaluation method for ubiquitous IoT systems in intelligent transportation of smart city},
  author={Chen, Zhicheng and Qu, Zhenzhe and Xiong, Nicholas and Liu, Anfeng and Dong, Mianxiong and Wang, Tian and Zhang, Shaobo},
  journal={IEEE Transactions on Intelligent Transportation Systems},
  volume={25},
  number={8},
  pages={9597--9607},
  year={2024},
  publisher={IEEE}
}

@article{sharma2022uav,
  title={UAV based long range environment monitoring system with Industry 5.0 perspectives for smart city infrastructure},
  author={Sharma, Rohit and Arya, Rajeev},
  journal={Computers \& Industrial Engineering},
  volume={168},
  pages={108066},
  year={2022},
  publisher={Elsevier}
}

@article{alqudsi2025uav,
  title={UAV swarms: research, challenges, and future directions},
  author={Alqudsi, Yunes and Makaraci, Murat},
  journal={Journal of Engineering and Applied Science},
  volume={72},
  number={1},
  pages={12},
  year={2025},
  publisher={Springer}
}

@inproceedings{zhao2024few,
  title={Few-shot learning and data augmentation for cross-domain uav fingerprinting},
  author={Zhao, Tianya and Wang, Ningning and Mao, Shiwen and Wang, Xuyu},
  booktitle={Proceedings of the 30th Annual International Conference on Mobile Computing and Networking},
  pages={2389--2394},
  year={2024}
}

@Article{5G-A_huawei,
    title = {Key Technologies for Low-Altitude Sensing in 5G-A Integrated Communication and Sensing Networks},
    journal = {JOURNAL OF SIGNAL PROCESSING},
    volume = {41},
    number = {5},
    pages = {787-806},
    year = {2025},
    issn = {1003-0530},
    author = {LIU Binyue and YANG Jianqiang and XU Bo and WANG Bolei and CAI Hua}
}

@misc{imt_2020,
    howpublished = {[Online]},
    note = {\url{https://www.cww.net.cn/article?id=600974}},
    title = {Research Report on Sensing Signal Processing Technologies},
    author = {IMT-2020(5G)},
    year = {2025}
}

@article{jain2005score,
  title={Score normalization in multimodal biometric systems},
  author={Jain, Anil and Nandakumar, Karthik and Ross, Arun},
  journal={Pattern recognition},
  volume={38},
  number={12},
  pages={2270--2285},
  year={2005},
  publisher={Elsevier}
}

@article{de2000mahalanobis,
  title={The mahalanobis distance},
  author={De Maesschalck, Roy and Jouan-Rimbaud, Delphine and Massart, D{\'e}sir{\'e} L},
  journal={Chemometrics and intelligent laboratory systems},
  volume={50},
  number={1},
  pages={1--18},
  year={2000},
  publisher={Elsevier}
}

@inproceedings{ester1996density,
  title={A density-based algorithm for discovering clusters in large spatial databases with noise},
  author={Ester, Martin and Kriegel, Hans-Peter and Sander, J{\"o}rg and Xu, Xiaowei and others},
  booktitle={kdd},
  volume={96},
  number={34},
  pages={226--231},
  year={1996}
}

@article{kuhn1955hungarian,
  title={The Hungarian method for the assignment problem},
  author={Kuhn, Harold W},
  journal={Naval research logistics quarterly},
  volume={2},
  number={1-2},
  pages={83--97},
  year={1955},
  publisher={Wiley Online Library}
}

@article{sutton1998reinforcement,
  title={Reinforcement learning: an introduction MIT Press},
  author={Sutton, Richard S and Barto, Andrew G},
  journal={Cambridge, MA},
  volume={22447},
  number={10},
  year={1998}
}

@inproceedings{duan2022pfilter,
  title={Pfilter: Building persistent maps through feature filtering for fast and accurate lidar-based slam},
  author={Duan, Yifan and Peng, Jie and Zhang, Yu and Ji, Jianmin and Zhang, Yanyong},
  booktitle={2022 IEEE/RSJ International Conference on Intelligent Robots and Systems (IROS)},
  pages={11087--11093},
  year={2022},
  organization={IEEE}
}

@article{nie20233d,
  title={3D object detection and tracking based on lidar-camera fusion and IMM-UKF algorithm towards highway driving},
  author={Nie, Chang and Ju, Zhiyang and Sun, Zhifeng and Zhang, Hui},
  journal={IEEE Transactions on Emerging Topics in Computational Intelligence},
  volume={7},
  number={4},
  pages={1242--1252},
  year={2023},
  publisher={IEEE}
}

@article{H-Ztest,
  title={A class of invariant consistent tests for multivariate normality},
  author={Henze, Norbert and Zirkler, Bernd},
  journal={Communications in statistics-Theory and Methods},
  volume={19},
  number={10},
  pages={3595--3617},
  year={1990},
  publisher={Taylor \& Francis}
}

@article{vaswani2017attention,
  title={Attention is all you need},
  author={Vaswani, Ashish and Shazeer, Noam and Parmar, Niki and Uszkoreit, Jakob and Jones, Llion and Gomez, Aidan N and Kaiser, {\L}ukasz and Polosukhin, Illia},
  journal={Advances in neural information processing systems},
  volume={30},
  year={2017}
}

@misc{core_network,
    howpublished = {[Online]},
    note = {\url{https://carrier.huawei.com/en/products/core-network}},
    title = {Cloud Core Network},
    author = {Huawei},
    year = {2025}
}

@techreport{Huawei2024CloudCore,
  author      = {Huawei},
  title       = {Intelligence + 5G-A: Elevating Connectivity
Beyond Boundaries},
  year        = {2024},
  url         = {https://www-file.huawei.com/admin/asset/v1/pro/view/d862bd7c8ce943d09e48dd56acd4cac5.pdf},
}

@misc{Mavic3,
    howpublished = {[Online]},
    note = {\url{https://enterprise.dji.com/mavic-3-enterprise}},
    title = {DJI Mavic 3 Enterprise Series},
    author = {DJI},
    year = {2022}
}

@article{wong2019reliable,
  title={Reliable accuracy estimates from k-fold cross validation},
  author={Wong, Tzu-Tsung and Yeh, Po-Yang},
  journal={IEEE Transactions on Knowledge and Data Engineering},
  volume={32},
  number={8},
  pages={1586--1594},
  year={2019},
  publisher={IEEE}
}

@misc{i5-11500,
    howpublished = {[Online]},
    year = {2021},
    note = {\url{https://www.intel.com/content/www/us/en/products/sku/212277/intel-core-i511500-processor-12m-cache-up-to-4-60-ghz/specifications.html}},
    author = {Intel},
    title = {i5-11500},
}

@article{qi2017pointnet++,
  title={Pointnet++: Deep hierarchical feature learning on point sets in a metric space},
  author={Qi, Charles Ruizhongtai and Yi, Li and Su, Hao and Guibas, Leonidas J},
  journal={Advances in neural information processing systems},
  volume={30},
  year={2017}
}

@inproceedings{kopp2023tackling,
  title={Tackling Clutter in Radar Data-Label Generation and Detection Using PointNet++},
  author={Kopp, Johannes and Kellner, Dominik and Piroli, Aldi and Dietmayer, Klaus},
  booktitle={2023 IEEE International Conference on Robotics and Automation (ICRA)},
  pages={1493--1499},
  year={2023},
  organization={IEEE}
}

@inproceedings{griebel2021anomaly,
  title={Anomaly detection in radar data using PointNets},
  author={Griebel, Thomas and Authaler, Dominik and Horn, Markus and Henning, Matti and Buchholz, Michael and Dietmayer, Klaus},
  booktitle={2021 IEEE International Intelligent Transportation Systems Conference (ITSC)},
  pages={2667--2673},
  year={2021},
  organization={IEEE}
}

@inproceedings{chu2025racformer,
  title={RaCFormer: Towards High-Quality 3D Object Detection via Query-based Radar-Camera Fusion},
  author={Chu, Xiaomeng and Deng, Jiajun and You, Guoliang and Duan, Yifan and Li, Houqiang and Zhang, Yanyong},
  booktitle={Proceedings of the Computer Vision and Pattern Recognition Conference},
  pages={17081--17091},
  year={2025}
}

@inproceedings{wang2023bi,
  title={Bi-lrfusion: Bi-directional lidar-radar fusion for 3d dynamic object detection},
  author={Wang, Yingjie and Deng, Jiajun and Li, Yao and Hu, Jinshui and Liu, Cong and Zhang, Yu and Ji, Jianmin and Ouyang, Wanli and Zhang, Yanyong},
  booktitle={Proceedings of the IEEE/CVF Conference on Computer Vision and Pattern Recognition},
  pages={13394--13403},
  year={2023}
}

@article{meng2024mmplace,
  title={mmplace: Robust place recognition with intermediate frequency signal of low-cost single-chip millimeter wave radar},
  author={Meng, Chengzhen and Duan, Yifan and He, Chenming and Wang, Dequan and Fan, Xiaoran and Zhang, Yanyong},
  journal={IEEE Robotics and Automation Letters},
  volume={9},
  number={6},
  pages={4878--4885},
  year={2024},
  publisher={IEEE}
}

@article{meng2025gr,
  title={Gr-fall: A fall detection system with gait recognition for indoor environments using siso mmwave radar},
  author={Meng, Chengzhen and He, Chenming and Wang, Dequan and Xiao, Yuxuan and Wang, Lingyu and Fan, Xiaoran and Zhang, Lu and Zhang, Yanyong},
  journal={Proceedings of the ACM on Interactive, Mobile, Wearable and Ubiquitous Technologies},
  volume={9},
  number={3},
  pages={1--26},
  year={2025},
  publisher={ACM New York, NY, USA}
}

@article{wang2024rdgait,
  title={Rdgait: A mmwave based gait user recognition system for complex indoor environments using single-chip radar},
  author={Wang, Dequan and Zhang, Xinran and Wang, Kai and Wang, Lingyu and Fan, Xiaoran and Zhang, Yanyong},
  journal={Proceedings of the ACM on Interactive, Mobile, Wearable and Ubiquitous Technologies},
  volume={8},
  number={3},
  pages={1--31},
  year={2024},
  publisher={ACM New York, NY, USA}
}

@inproceedings{he2024see,
  title={See Through Vehicles: Fully Occluded Vehicle Detection with Millimeter Wave Radar},
  author={He, Chenming and Meng, Chengzhen and He, Chunwang and Fan, Xiaoran and Wang, Beibei and Yan, Yubo and Zhang, Yanyong},
  booktitle={Proceedings of the 30th Annual International Conference on Mobile Computing and Networking},
  pages={740--754},
  year={2024}
}

@inproceedings{he2025ghost,
  title={Ghost Points Matter: Far-Range Vehicle Detection with a Single mmWave Radar in Tunnel},
  author={He, Chenming and Xia, Rui and Meng, Chengzhen and Fan, Xiaoran and Wang, Dequan and Ren, Haojie and Ji, Jianmin and Zhang, Yanyong},
  booktitle={Proceedings of the 31st Annual International Conference on Mobile Computing and Networking},
  pages={650--666},
  year={2025}
}

@article{wu2023precise,
  title={Precise UAV MMW-vision positioning: A modal-oriented self-tuning fusion framework},
  author={Wu, Guangyu and Zhou, Fuhui and Meng, Chengzhen and Li, Xiang-Yang},
  journal={IEEE Journal on Selected Areas in Communications},
  volume={42},
  number={1},
  pages={6--20},
  year={2023},
  publisher={IEEE}
}

@article{wu2024vehicle,
  title={A vehicle-mounted radar-vision system for precisely positioning clustering uavs},
  author={Wu, Guangyu and Zhou, Fuhui and Wong, Kai Kit and Li, Xiang-Yang},
  journal={IEEE Journal on Selected Areas in Communications},
  volume={42},
  number={10},
  pages={2688--2703},
  year={2024},
  publisher={IEEE}
}

@article{wang2024survey,
  title={A survey on vision-based anti unmanned aerial vehicles methods},
  author={Wang, Bingshu and Li, Qiang and Mao, Qianchen and Wang, Jinbao and Chen, CL Philip and Shangguan, Aihong and Zhang, Haosu},
  journal={Drones},
  volume={8},
  number={9},
  pages={518},
  year={2024},
  publisher={MDPI}
}

@inproceedings{iizuka2023millisign,
  title={Millisign: mmwave-based passive signs for guiding uavs in poor visibility conditions},
  author={Iizuka, Tatsuya and Sasatani, Takuya and Nakamura, Toru and Kosaka, Naoko and Hisada, Masaki and Kawahara, Yoshihiro},
  booktitle={Proceedings of the 29th Annual International Conference on Mobile Computing and Networking},
  pages={1--15},
  year={2023}
}

@inproceedings{zhang2023mmhawkeye,
  title={mmHawkeye: Passive UAV detection with a COTS mmWave radar},
  author={Zhang, Jia and Na, Xin and Xi, Rui and Sun, Yimiao and He, Yuan},
  booktitle={2023 20th Annual IEEE International Conference on Sensing, Communication, and Networking (SECON)},
  pages={267--275},
  year={2023},
  organization={IEEE}
}

@inproceedings{lu2024integrated,
  title={Integrated sensing and communication enabled multiple base stations cooperative UAV detection},
  author={Lu, Xi and Wei, Zhiqing and Xu, Ruizhong and Wang, Lin and Lu, Bohao and Piao, Jinghui},
  booktitle={2024 IEEE International Conference on Communications Workshops (ICC Workshops)},
  pages={1882--1887},
  year={2024},
  organization={IEEE}
}

@article{wu2021comprehensive,
  title={A comprehensive overview on 5G-and-beyond networks with UAVs: From communications to sensing and intelligence},
  author={Wu, Qingqing and Xu, Jie and Zeng, Yong and Ng, Derrick Wing Kwan and Al-Dhahir, Naofal and Schober, Robert and Swindlehurst, A Lee},
  journal={IEEE Journal on Selected Areas in Communications},
  volume={39},
  number={10},
  pages={2912--2945},
  year={2021},
  publisher={IEEE}
}

@inproceedings{lin2024integrated,
  title={Integrated Sensing and Communications: A Survey of Recent Progress Toward 5G-A and 6G},
  author={Lin, Xu and Du, Zhaopeng and Mei, Lin and Zhang, Ruoyu},
  booktitle={International Conference on Wireless and Satellite Systems},
  pages={245--262},
  year={2024},
  organization={Springer}
}

@article{rezaei2015robust,
  title={Robust vehicle detection and distance estimation under challenging lighting conditions},
  author={Rezaei, Mahdi and Terauchi, Mutsuhiro and Klette, Reinhard},
  journal={IEEE transactions on intelligent transportation systems},
  volume={16},
  number={5},
  pages={2723--2743},
  year={2015},
  publisher={IEEE}
}

@inproceedings{poitevin2017challenges,
  title={Challenges in detecting UAS with radar},
  author={Poitevin, Pierre and Pelletier, Michel and Lamontagne, Patrick},
  booktitle={2017 International Carnahan Conference on Security Technology (ICCST)},
  pages={1--6},
  year={2017},
  organization={IEEE}
}

@article{yang2025hierarchical,
  title={Hierarchical Reinforcement Learning-Based Beam Selection for Integrated Sensing and Communication Systems},
  author={Yang, Ruming and Li, Xingkang and Huang, Yongming and Yang, Luxi and Zhang, Wei},
  journal={IEEE Transactions on Wireless Communications},
  year={2025},
  publisher={IEEE}
}

@article{saikia2024hybrid,
  title={Hybrid deep reinforcement learning for enhancing localization and communication efficiency in RIS-aided cooperative ISAC systems},
  author={Saikia, Prajwalita and Singh, Keshav and Huang, Wan-Jen and Duong, Trung Q},
  journal={IEEE internet of things journal},
  volume={11},
  number={18},
  pages={29494--29510},
  year={2024},
  publisher={IEEE}
}

@article{hu2024joint,
  title={Joint range-velocity-azimuth estimation for OFDM-based integrated sensing and communication},
  author={Hu, Zelin and Ye, Qibin and Huang, Yixuan and Hu, Su and Yang, Gang},
  journal={IEEE Transactions on Wireless Communications},
  volume={23},
  number={10},
  pages={12933--12948},
  year={2024},
  publisher={IEEE}
}

@article{lu2024deep,
  title={Deep learning based multi-node ISAC 4D environmental reconstruction with uplink-downlink cooperation},
  author={Lu, Bohao and Wei, Zhiqing and Wu, Huici and Zeng, Xinrui and Wang, Lin and Lu, Xi and Mei, Dongyang and Feng, Zhiyong},
  journal={IEEE Internet of Things Journal},
  year={2024},
  publisher={IEEE}
}

@article{motie2024self,
  title={Self UAV localization using multiple base stations based on TDoA measurements},
  author={Motie, Samaneh and Zayyani, Hadi and Salman, Mohammad and Bekrani, Mehdi},
  journal={IEEE Wireless Communications Letters},
  volume={13},
  number={9},
  pages={2432--2436},
  year={2024},
  publisher={IEEE}
}

@article{wang2019assistant,
  title={Assistant vehicle localization based on three collaborative base stations via SBL-based robust DOA estimation},
  author={Wang, Huafei and Wan, Liangtian and Dong, Mianxiong and Ota, Kaoru and Wang, Xianpeng},
  journal={IEEE Internet of Things Journal},
  volume={6},
  number={3},
  pages={5766--5777},
  year={2019},
  publisher={IEEE}
}

@inproceedings{lu2024intergrated,
  title={Intergrated Sensing and Communications (ISAC) in 5G-Advanced for High-Precison Localization and Tracking of Vessels at Sea},
  author={Lu, Fangkai and Huang, Zhen and Li, Yongqiang and Song, Xiaofeng and Shi, Xutao and Wang, Yanfei},
  booktitle={2024 10th International Conference on Computer and Communications (ICCC)},
  pages={1288--1292},
  year={2024},
  organization={IEEE}
}

@article{liu2025key,
  title={Key Technologies for Low-Altitude Sensing in 5G-A Integrated Communication and Sensing Networks},
  author={Liu, Binyue and Yang, Jianqiang and Xu, Bo and Wang, Bolei and Cai, Hua},
  journal={Journal of Signal Processing},
  volume={41},
  number={5},
  pages={787--806},
  year={2025}
}

@inproceedings{cheng2025convtimenet,
  title={Convtimenet: A deep hierarchical fully convolutional model for multivariate time series analysis},
  author={Cheng, Mingyue and Yang, Jiqian and Pan, Tingyue and Liu, Qi and Li, Zhi and Wang, Shijin},
  booktitle={Companion Proceedings of the ACM on Web Conference 2025},
  pages={171--180},
  year={2025}
}

\appendix

\section{Cube Size for Space Partitioning}
\label{appendix-b}

\begin{figure}[h]
    \centering
    \setlength{\abovecaptionskip}{2mm}
    \includegraphics[width=0.95\linewidth]{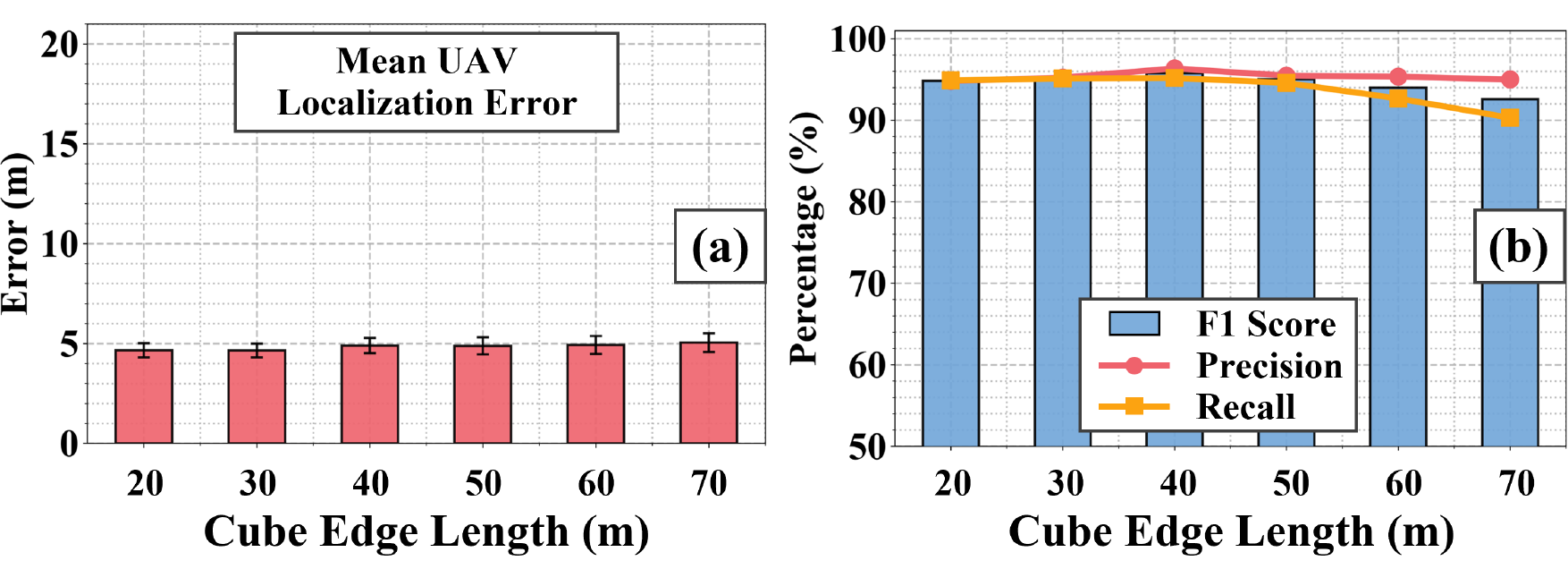}
    \caption{UAV tracking performance across different cube edge lengths: (a) mean localization error; (b) F1 score.}
    \label{fig: result_pnt_voxel}
    \vspace{-2mm}
\end{figure}

We partition the 3D space into cubes and model the fingerprint of noise points within each cube using a multivariate Gaussian, as described in Section~\ref{sec: gaussian_model}.
In theory, the cube size should not be set too large; otherwise, a single cube may encompass multiple reflectors with significantly different scattering properties, causing the statistical characteristics of the noise points to deviate from a Gaussian distribution.

Therefore, we vary the cube sizes and evaluate the resulting UAV tracking performance on our dataset, as illustrated in Figure~\ref{fig: result_pnt_voxel}.
We find that when the cube edge length is kept within 50 m, the system performance remains stable, with the F1 score consistently above 95\%. However, once the edge length exceeds 60 m, the performance begins to degrade, primarily reflected in a noticeable drop in recall.

This trend is expected: larger cubes are more likely to mix reflections from different types of surfaces, reducing the local statistical consistency required for reliable Gaussian modeling. As the fitted distribution becomes less representative, more true UAV points may be incorrectly filtered out, leading to reduced recall. In summary, we recommend keeping the cube edge length below 50~m when partitioning the space. In our system, we ultimately set it to 40~m.

\end{document}